\documentclass[aps, pre, twocolumn, superscriptaddress, preprintnumbers]{revtex4-2}
\usepackage{bm, amsmath, amsfonts, amssymb, ascmac, mathtools, braket}
\usepackage{times}
\usepackage{multirow}
\usepackage{graphicx}
\usepackage{float, color, xcolor}

\usepackage[whole]{bxcjkjatype} 
\usepackage{subfigure} 
\usepackage{amscd} 

\usepackage{bbm}
\usepackage{tabularx}

\usepackage{comment}

\definecolor{rred}{rgb}{0.8, 0.0, 0.0}
\definecolor{bblue}{rgb}{0.0, 0.0, 0.8}

\usepackage[
pagebackref=false,
colorlinks=true,
linkcolor=bblue,
urlcolor=bblue,
filecolor=black,
citecolor=rred,
pdfstartview=FitV,
pdftitle={},
pdfauthor={},
pdfsubject={},
pdfkeywords={},
pdfpagemode=None,
bookmarksopen=true
]{hyperref}

\newcommand{\ii}{\text{i}}
\newcommand{\Heav}{H}

\begin{document}

\title{Exact probability distributions of complex spacing ratios in non-Hermitian random matrices}

\author{Kohei Kawabata}
\email{kawabata@issp.u-tokyo.ac.jp}
\affiliation{Institute for Solid State Physics, University of Tokyo, Kashiwa, Chiba 277-8581, Japan}

\date{\today}

\begin{abstract}
The complex spacing ratio is the complex displacement from a reference eigenvalue to its nearest neighbor divided by the corresponding displacement to its next-to-nearest neighbor.
Its statistics provide a useful diagnostic of spectral correlations and nonintegrability in open quantum systems.
Here, starting from the exact joint eigenvalue probability densities,
we derive finite-$N$ complex-spacing-ratio distributions for the Gaussian ensembles of non-Hermitian random matrices in classes AI$^{\dag}$ and AII$^{\dag}$,
realized by complex symmetric and complex self-dual random matrices, respectively.
In class AII$^{\dag}$, we obtain an exact algebraic expression for arbitrary $N$ and explicitly evaluate the distributions and representative moments for $N=3, 4, 5, 6$.
In class AI$^{\dag}$, 
although the joint density retains a noncompact integral over nonunitary eigenvector degrees of freedom, 
we analytically derive a normalized one-dimensional integral representation for $N=3$ and determine the asymptotic behavior, 
including a logarithmic correction to the cubic level repulsion and a nonanalytic contribution to the angular density.
We further confirm these analytical results through direct numerical diagonalization of non-Hermitian random matrices.
\end{abstract}

\maketitle

\section{Introduction}

Hermitian random matrices provide a statistical description of energy spectra in complicated quantum systems~\cite{Mehta-textbook, Forrester-textbook}. 
In Wigner's original application to highly excited nuclei,
an intractable microscopic Hamiltonian was replaced by a statistical ensemble to capture correlations among energy levels rather than individual levels themselves~\cite{Wigner-51, Wigner-58}. 
This perspective subsequently became central to quantum chaos, 
where generic chaotic systems exhibit the spectral correlations of Hermitian random matrices~\cite{BGS-84, Haake-textbook}. 
Whereas the mean density of states can depend sensitively on microscopic details, 
local multi-point spectral correlations become universal.
Such universal bulk spectral correlations are organized by time-reversal symmetry according to Dyson's threefold way~\cite{Dyson-62}: 
class A in the absence of time-reversal symmetry, 
and classes AI and AII in the presence of time-reversal symmetry with positive and negative signs, respectively. 
A standard manifestation of the Wigner-Dyson statistics is level repulsion, 
characterized by the distribution of spacings between neighboring eigenvalues. 
Comparing spacing distributions across different systems generally requires spectral unfolding to remove variations in the local mean density. 
Ratios of consecutive spacings circumvent such an explicit unfolding procedure by canceling the common local spectral scale~\cite{Oganesyan-07, Atas-13}. 
Consequently, the spacing-ratio statistics have become widely used as a diagnostic of quantum chaos and nonintegrability.

Non-Hermitian random matrices extend this statistical approach to open quantum systems~\cite{Ginibre-65, Girko-85, Sommers-88, Grobe-89, Lehmann-91, Feinberg-97, Chalker-98, Nishigaki-02, Kanzieper-05, Forrester-07, Akemann-09, Byun-Forrester-textbook}. 
In open quantum systems, dynamical (super)operators,
such as Lindbladians and Kraus maps, 
are generally non-Hermitian~\cite{Nielsen-textbook, Breuer-textbook, Rivas-textbook}. 
As in the Hermitian setting,
the local correlations of their complex spectra have been used to characterize dissipative quantum chaos and nonintegrability~\cite{Grobe-88, Haake-92, Xu-19, Hamazaki-19, Denisov-19, Can-19JPhysA, Sa-20JPhysA, Akemann-19, Wang-20, JiachenLi-21, GarciaGarcia-22PRX, Prasad-22, GJ-23, Ghosh-22, Sa-23, Kawabata-23, Sa-Ribeiro-Denisov-26review}.
An important feature of non-Hermitian systems is the enriched symmetry classification~\cite{Bernard-LeClair-02, KSUS-19, Hamazaki-20, Kawabata-Ryu-26review}.
Specifically, complex conjugation and transposition are no longer equivalent, 
giving rise to two distinct extensions of time-reversal symmetry~\cite{Bernard-LeClair-02, KSUS-19}.
On the one hand, time-reversal symmetry involves complex conjugation and underlies Ginibre's threefold classification (classes A, AI, and AII)~\cite{Ginibre-65}.
Although it relates eigenvalues $z$ and their complex conjugates $z^*$,
it does not distinguish the generic bulk statistics of the three Ginibre ensembles away from the real axis in the large-matrix limit. 
On the other hand, time-reversal symmetry$^{\dag}$, 
defined through transposition,
constrains left and right eigenvectors for the same eigenvalue and changes local spectral correlations throughout the complex plane~\cite{Hamazaki-20}. 
Its absence, and its presence with two possible signs yield classes A, AI$^{\dag}$, and AII$^{\dag}$~\cite{KSUS-19},
represented by generic complex, complex symmetric, and complex self-dual matrices, respectively. 
The three distinct bulk statistics constitute a non-Hermitian counterpart of Dyson's threefold way.

The distinct universal spectral statistics in the two classes with time-reversal symmetry$^{\dag}$ (i.e., classes AI$^{\dag}$ and AII$^{\dag}$) were first identified numerically~\cite{Hamazaki-20},
and their analytical characterization has remained incomplete, 
despite several advances~\cite{Akemann-22, Akemann-25, Kulkarni-25, Forrester-25, Chen-25, Akemann-Fyodorov-Savin-25}.
Remarkably, a recent work has obtained exact joint eigenvalue probability distributions for the Gaussian ensembles in classes AI$^{\dag}$ and AII$^{\dag}$,
by relating them to scattering states of the Calogero model~\cite{Xiao-26}.
Whereas the probability density for class AII$^{\dag}$ is described by a finite polynomial,
that for class AI$^{\dag}$ retains a noncompact integral over eigenvector degrees of freedom. 
Importantly, both involve irreducible many-eigenvalue interactions,
in contrast to the purely pairwise Coulomb-gas structure of the standard Ginibre statistics. 
These results have also enabled exact finite-$N$ calculations of complex level-spacing distributions, 
where $N$ denotes the number of different complex eigenvalues,
and provide a starting point for deriving other spectral observables.
Complementary recent approaches based on characteristic polynomials and replica nonlinear sigma models have also yielded universal eigenvalue and eigenvector statistics~\cite{Chen-26, Fyodorov-26, Chen-26vec}.

As in real spectra for Hermitian matrices, 
complex eigenvalue spacings depend on the local eigenvalue density. 
Unfolding is more involved in the complex plane because spatial variations of a two-dimensional density must be removed.
To circumvent this unfolding procedure, 
a complex analog of spacing ratios was introduced,
in which the complex displacement from a reference eigenvalue to its nearest neighbor is divided by that to its next-to-nearest neighbor, 
with both neighbors selected according to Euclidean distance~\cite{Sa-20}.  
Both modulus and argument of this complex spacing ratio encode relevant information about spectral correlations. 
For uncorrelated eigenvalues, 
the ratio is uniformly distributed over the unit disk. 
By contrast, random-matrix correlations suppress small moduli and generate nontrivial angular dependence, 
with small relative angles depleted in the bulk. 
This angular structure provides information absent from the distribution of the ratio modulus alone.

Despite their utility, 
analytical results for complex-spacing-ratio distributions remain limited.
Whereas the Wigner-like surmises derived from small Hermitian random matrices provide accurate approximations to the large-$N$ real-spacing-ratio distributions,
the complex-spacing-ratio distributions of small non-Hermitian random matrices,
such as those for $N=3$,
substantially deviate from their large-$N$ counterparts.
Accordingly, accurately capturing the finite-$N$ effects is particularly important for non-Hermitian random matrices.
For class A, finite-$N$ algebraic expressions have been derived as exact distributions conditioned on the presence  of a reference eigenvalue at the origin~\cite{Dusa-22, Akemann-26}. 
These finite-$N$ formulas also provide accurate approximations to the universal bulk statistics as the matrix dimension $N$ increases. 
However, comparable analytical distributions for classes AI$^{\dag}$ and AII$^{\dag}$ have hitherto remained unavailable,
even for small non-Hermitian random matrices.
In this respect, it is also noteworthy that the complex-spacing-ratio statistics encode rich ordering information, 
requiring both the nearest- and next-to-nearest-neighbor constraints, 
and thus the full hierarchy of multi-point eigenvalue correlations, 
beyond two-point or three-point correlations.

In this work, we derive finite-$N$ complex-spacing-ratio distributions in classes AI$^{\dag}$ and AII$^{\dag}$,
from the exact joint eigenvalue densities reported in Ref.~\cite{Xiao-26}. 
Following the strategy for class A~\cite{Dusa-22, Akemann-26},
we condition on the presence of a reference eigenvalue at the spectral origin and retain the neighbor-ordering constraints.
For class AII$^{\dag}$, we derive an exact algebraic expression for an arbitrary number $N$ of different eigenvalues and explicitly evaluate the distributions and representative moments for $N=3,4,5,6$. 
We confirm these analytical results through comparisons with independent numerical diagonalization of non-Hermitian random matrices. 
The resulting distributions resolve symmetry-dependent radial and angular correlations, 
and comparison with larger matrices reveals nonmonotonic finite-size behavior.
For class AI$^{\dag}$, the remaining noncompact orbital integral makes calculations more challenging. 
For $N=3$, we derive a normalized one-dimensional integral representation involving the complete elliptic integral of the first kind.
From this expression, we analytically determine the small-modulus and small-argument asymptotics, 
including a logarithmic correction to the cubic level repulsion and a nonanalytic contribution to the angular density. 
For generic $N$, we evaluate the finite-size statistics by Monte Carlo integration of the exact joint eigenvalue densities, 
without generating and diagonalizing random matrices.
In addition to the results for class A~\cite{Dusa-22, Akemann-26},
these results provide finite-size benchmarks for complex-spacing-ratio distributions across the non-Hermitian threefold way.

The remainder of this work is organized as follows. 
In Sec.~\ref{sec:RMT}, we introduce non-Hermitian random matrices and their time-reversal symmetry$^{\dag}$. 
In Sec.~\ref{sec:CSR}, we define complex spacing ratios and formulate their origin-conditioned distribution. 
In Sec.~\ref{sec:A}, we review the complex-spacing-ratio distribution in class A.
In Secs.~\ref{sec:AII} and \ref{sec:AI}, we present the corresponding distributions for classes AII$^{\dag}$ and AI$^{\dag}$, respectively. 
In Sec.~\ref{sec:conclusion}, we conclude this work. 
In Appendix~\ref{appendix:origin}, we explain a numerical procedure for evaluating the origin-conditioned complex-spacing-ratio distributions.
In Appendix~\ref{appendix:A1}, we provide a general finite-$N$ integral representation for class AI$^{\dag}$.

\section{Threefold way of non-Hermitian random matrices}
    \label{sec:RMT}

In non-Hermitian systems, the symmetry relevant to local complex-eigenvalue correlations is time-reversal symmetry$^{\dag}$~\cite{KSUS-19},
\begin{align}
\mathcal{T}H^T\mathcal{T}^{-1}=H,
\qquad
\mathcal{T}\mathcal{T}^*=\pm1,
\end{align}
with a unitary matrix $\mathcal{T}$ (i.e., $\mathcal{T}\mathcal{T}^{\dag} = \mathcal{T}^{\dag} \mathcal{T} = 1$).
Replacing $H^T$ by $H^*$ gives time-reversal symmetry (i.e., $\mathcal{T}H^*\mathcal{T}^{-1}=H$).
These two definitions coincide for Hermitian matrices but are distinct for non-Hermitian matrices.
Time-reversal symmetry pairs eigenvalues $z$ and $z^*$, 
and consequently,
the real and quaternion-real Ginibre ensembles share the same complex Ginibre bulk statistics away from the real axis in the large-$N$ limit~\cite{Ginibre-65, Byun-Forrester-textbook}.
By contrast, time-reversal symmetry$^{\dag}$ relates left and right eigenvectors for the same eigenvalue and influences local correlations throughout the complex spectrum~\cite{Hamazaki-20}.
Its absence, and its presence with signs $+1$ and $-1$ distinguish classes A, AI$^{\dag}$, and AII$^{\dag}$, respectively.

Class A consists of unrestricted complex matrices and corresponds to the complex Ginibre ensemble.
In class AI$^{\dag}$, we can choose a suitable basis in which we have $\mathcal{T}=I_N$, 
so that $H^T=H$, i.e., $H$ is complex symmetric.
In class AII$^{\dag}$, we choose
\begin{equation}
\mathcal{T}=
\begin{pmatrix}
0&I_N\\
-I_N&0
\end{pmatrix},
\end{equation}
so that $H$ is complex self-dual.
Each complex eigenvalue is then generally twofold degenerate (i.e., Kramers degeneracy).
Throughout this work, $N$ denotes the number of different eigenvalues, 
and the matrix dimension is $N$ in classes A and AI$^{\dag}$, and $2N$ in class AII$^{\dag}$.
Each degenerate pair is counted once before identifying neighboring eigenvalues.

We consider the Gaussian ensembles of non-Hermitian random matrices $H$ with probability measures proportional to
\begin{equation}
\begin{cases}
e^{-\operatorname{Tr}\,(H^\dagger H)}\,dH
& ( \text{classes A and AI}^{\dag} ),\\
e^{-\operatorname{Tr}\,(H^\dagger H)/2}\,dH
& ( \text{class AII}^{\dag} ).
\end{cases}
\end{equation}
Here, $dH$ denotes the flat measure over the independent real and imaginary matrix components subject to the corresponding symmetry constraints.
All the three Gaussian measures are invariant under $H\mapsto e^{\ii\phi}H$ ($\phi \in \mathbb{R}$), 
implying invariance of the joint eigenvalue density under a common rotation of the spectrum.
Let $\rho_N \left( z_1,\ldots,z_N \right)$ be the permutation-symmetric joint density of the $N$ different eigenvalues, 
normalized by
\begin{equation}    
    \int \left( \prod_{n=1}^{N}d^2z_n \right) \rho_N \left( z_1,\ldots,z_N \right) = 1.
\end{equation}

These three classes of bulk spectral statistics were first identified numerically~\cite{Hamazaki-20}.
A recent analytical work~\cite{Xiao-26} has obtained exact finite-$N$ joint eigenvalue probability densities $\rho_N \left( z_1,\ldots,z_N \right)$ for classes AI$^{\dag}$ and AII$^{\dag}$.
Unlike the complex Ginibre counterpart,
these joint densities cannot be represented using pairwise eigenvalue interactions alone.
In the following, we use these joint eigenvalue densities to derive complex-spacing-ratio distributions.
It should be noted that the finite-$N$ formulas below apply to the Gaussian ensembles defined above and need not coincide with the universal large-$N$ bulk distributions.

\section{Complex spacing ratio}
    \label{sec:CSR}

Suppose that a complex eigenvalue $z \in \mathbb{C}$ is chosen as a reference point,
and let $z_{\rm NN} \in \mathbb{C}$ and $z_{\rm NNN} \in \mathbb{C}$ denote its nearest-neighbor and next-to-nearest-neighbor eigenvalues, respectively.
Then, we define the complex spacing ratio as~\cite{Sa-20}
\begin{equation}
    \eta \coloneqq \frac{z_{\rm NN} - z}{z_{\rm NNN} - z} \eqqcolon re^{\ii\theta} \in \mathbb{C}.
        \label{eq:CSR}
\end{equation}
Here, $\left| \eta \right| = r \in \left[ 0, 1 \right]$ is the ratio of the distances $\left| z_{\rm NN} - z \right|$ and $\left| z_{\rm NNN} - z \right|$ in the complex plane,
while $\arg \eta = \theta \in \left( -\pi, \pi \right]$ is the relative angle of $z_{\rm NN}$ and $z_{\rm NNN}$ with respect to $z$.
In particular, $\theta \simeq 0$ means that $z_{\rm NN}$ and $z_{\rm NNN}$ lie in nearly the same direction from $z$,
and $\theta \simeq \pi$ means that they lie in nearly the opposite directions.
In this manner, the complex spacing ratio captures the local geometry of three eigenvalues in the complex plane.
Nevertheless, it should also be noted that the statistics of complex spacing ratios cannot be reduced solely to the three-point correlation function of complex eigenvalues.
Indeed, the requirement that $z_{\rm NN}$ and $z_{\rm NNN}$ be the nearest and next-to-nearest neighbors of $z$ imposes the additional condition $\left| z_n - z \right| > \left| z_{\rm NNN} - z \right|$ for all the eigenvalues $z_n$'s other than $z$, $z_{\rm NN}$, and $z_{\rm NNN}$.

Now, suppose that we have $N$ complex eigenvalues $z_1, z_2, \cdots, z_N \in \mathbb{C}$ whose statistics are described by the joint eigenvalue density function $\rho_N \left( z_1, z_2, \cdots, z_N \right)$.
Without loss of generality, we can choose $z_1$ as a reference eigenvalue and label its nearest and next-to-nearest neighbors as $z_2$ and $z_3$, respectively,
satisfying
\begin{equation}
    \left| z_2 - z_1 \right| < \left| z_3 - z_1 \right| < \left| z_n - z_1 \right|~~~~\left( n = 4, 5, \cdots, N \right).
        \label{eq:eig-ordering}
\end{equation}
Then, the complex spacing ratio $\eta$ in Eq.~\eqref{eq:CSR} becomes
\begin{equation}
    \eta = \frac{z_2 - z_1}{z_3 - z_1}.
        \label{eq:CSR-v2}
\end{equation}
Accordingly, the probability density function of $\eta$ is given as
\begin{align}
    &p_N \left( \eta \right) \propto \int d^2 z_1 \cdots d^2 z_N~\rho_N \left( z_1, \cdots, z_N \right) \nonumber \\
    &\quad\times\delta^{(2)} \left( \eta - \frac{z_2-z_1}{z_3-z_1} \right) \Heav \left( \left| z_3 - z_1 \right| - \left| z_2 - z_1 \right| \right) \nonumber \\
    &\quad\times\prod_{n=4}^{N} \Heav \left( \left| z_n - z_1\right| - \left| z_3 - z_1 \right| \right),
\end{align}
with the Heaviside step function $H$.
The radial and angular marginal densities are respectively given as
\begin{align}
    p_{r, N} \left( r \right) &= r \int_{-\pi}^{\pi} d\theta\,p_N\,( re^{\ii \theta} ), \\
    p_{\theta, N} \left( \theta \right) &= \int_{0}^{1} dr\,r\,p_N\,( re^{\ii \theta} ).
        \label{eq:marginal_theta}
\end{align}
For the Gaussian ensembles considered here, 
invariance under
complex conjugation $H \mapsto H^*$ maps $\eta$ to $\eta^*$ while preserving the
neighbor ordering. 
Hence, the angular density $p_{\theta, N} \left( \theta \right)$ is an even function and expanded as
\begin{equation}
    p_{\theta, N} \left( \theta \right) = \frac{1}{2\pi} \left( 1 + 2 \sum_{n=1}^{\infty} \braket{\cos n\theta} \cos n\theta \right).
\end{equation}
Thus, the angular moments $\braket{\cos n\theta}$ constitute the Fourier coefficients of $p_{\theta, N} \left( \theta \right)$.

\subsection{Origin-conditioned distribution}

To obtain the statistics of complex spacing ratios, 
it is convenient to choose the reference eigenvalue $z_1$ at the origin:
$z_1 = 0$.
Then, Eq.~\eqref{eq:eig-ordering} reduces to
\begin{equation}
    \left| z_2 \right| < \left| z_3 \right| < \left| z_n \right| \quad \left( n = 4, 5, \cdots, N \right),
\end{equation}
and the complex spacing ratio $\eta$ in Eq.~\eqref{eq:CSR-v2} reduces to 
\begin{equation}
    \eta = \frac{z_2}{z_3},
\end{equation}
which simplifies the problem.
Since the spectral origin is a regular bulk point of the ensembles considered here, 
the origin-conditioned statistics are expected to approach the corresponding universal bulk statistics in the large-$N$ limit.
However, these statistics do not necessarily describe the statistics for a generic reference eigenvalue at finite $N$.
In other words, we here consider the eigenvalue statistics conditioned on the presence of an eigenvalue at the origin.
More explicitly, given the joint eigenvalue density $\rho_N \left( z_1, z_2, \cdots, z_N \right)$,
the origin-conditioned density $\rho_N^{(0)} \left( z_2, z_3, \cdots, z_N \right)$ is given by
\begin{equation}
    \rho_N^{(0)} \left( z_2, z_3, \cdots, z_N \right) \propto \rho_N \left( 0, z_2, \cdots, z_N \right).
\end{equation}

Then, the probability density of the complex spacing ratio $\eta = z_2/z_3$ is given as
\begin{align}
    &p_N^{(0)} \left( \eta \right) \propto \int d^2 z_2 \cdots d^2 z_N~\rho_N^{(0)} \left( z_2, \cdots, z_N \right) \nonumber \\
    &\times\delta^{(2)} \left( \eta - \frac{z_2}{z_3} \right) \Heav \left( \left| z_3 \right| - \left| z_2 \right| \right) \prod_{n=4}^{N} \Heav \left( \left| z_n\right| - \left| z_3 \right| \right).
        \label{eq:origin-conditioned}
\end{align}
Using the identity
\begin{equation}
    \delta^{(2)} \left( \eta - \frac{z_2}{z_3} \right) = \left| z_3 \right|^2 \delta^{(2)} \left( z_2 - \eta z_3 \right),
\end{equation}
we perform the integral over $z_2$ to obtain
\begin{align}
    &p_N^{(0)} \left( \eta \right) \propto 
    \int d^2 z_3 \cdots d^2 z_N \left| z_3 \right|^2 \nonumber \\
    &\quad\times \rho_N^{(0)} \left( \eta z_3, z_3, \cdots, z_N \right) \prod_{n=4}^{N} \Heav \left( \left| z_n\right| - \left| z_3 \right| \right).
        \label{eq:pN}
\end{align}
Hereafter, we assume $\left| \eta \right| \leq 1$.
We next parametrize $z_3$ as 
\begin{equation}
    z_3 \eqqcolon \sqrt{t} e^{\ii \phi}.
\end{equation}
Employing rotational invariance to perform the integral over $\phi$,
we obtain
\begin{align}
    &p_N^{(0)} \left( \eta \right) \propto 
    \int_0^{\infty} dt\,t
    \int d^2 z_4 \cdots d^2 z_N \nonumber \\
    &\quad\times \rho_N^{(0)}\,( \eta \sqrt{t}, \sqrt{t}, z_4, \cdots, z_N ) \prod_{n=4}^{N} \Heav\,( \left| z_n\right|^2 - t ).
        \label{eq:pN-t}
\end{align}
This expression applies to all the symmetry classes investigated below.

\section{Class A}
    \label{sec:A}

For the Gaussian ensemble in class A, the joint eigenvalue probability density function reads~\cite{Byun-Forrester-textbook}
\begin{equation}
    \rho_N \left( z_1, \cdots, z_N \right) \propto e^{- \sum_{n=1}^{N} \left| z_n \right|^2} \left| \Delta_N \left( z_1, \cdots, z_N \right) \right|^2
        \label{eq:A-jpdf}
\end{equation}
with the Vandermonde determinant
\begin{equation}    
    \Delta_N \left( z_1, \cdots, z_N \right) = \prod_{i<j} \left( z_i - z_j \right).
        \label{eq:Vandermonde}
\end{equation}
In the following, we review derivations of the complex-spacing-ratio distribution $p_{N}^{(0)} \left( \eta \right)$~\cite{Dusa-22, Akemann-26},
beginning with the simplest cases of $N = 3, 4$ and then extending the analysis to arbitrary $N$.

\subsection{\texorpdfstring{$N=3$}{N=3}}

Let us evaluate the complex-spacing-ratio distribution $p_N^{(0)} \left( \eta \right)$ for the simplest case with $N=3$.
For $N=3$, Eq.~\eqref{eq:A-jpdf} reduces to
\begin{equation}
    \rho_3^{(0)} \left( z_2, z_3 \right) \propto \left| z_2 \right|^2 \left| z_3 \right|^2 \left| z_2 - z_3 \right|^2 e^{- |z_2|^2 - |z_3|^2},
\end{equation}
and Eq.~\eqref{eq:pN} yields
\begin{align}
    &p_3^{(0)} \left( \eta \right) \propto \int d^2 z_3 \left| z_3 \right|^2 \rho_3^{(0)} \left( \eta z_3, z_3 \right) \nonumber \\
    &\propto \left| \eta \right|^2 \left| 1-\eta \right|^2 \int d^2 z_3 \left| z_3 \right|^8 e^{- (1+|\eta|^2)\,|z_3|^2} \nonumber \\
    &\propto \frac{\left| \eta \right|^2 \left| 1-\eta \right|^2}{( 1 + \left| \eta \right|^2 )^5}.
\end{align}
After imposing the normalization condition, we obtain
\begin{equation}
    p_3^{(0)} \left( \eta \right) = \frac{12}{\pi} \frac{\left| \eta \right|^2 \left| 1-\eta \right|^2}{( 1 + \left| \eta \right|^2 )^5},
        \label{eq:p-N3}
\end{equation}
with the marginal densities
\begin{align}
    p_{r, 3} \left( r \right) &= \frac{24r^3}{\left( 1+r^2 \right)^4}, \label{eq:p_r_N3_A} \\
    p_{\theta, 3} \left( \theta \right) &= \frac{1}{2\pi} \left( 1 - \frac{9\pi}{32} \cos \theta \right).
        \label{eq:p_theta_N3_A}
\end{align}
The corresponding moments are
\begin{align}
    &\braket{r} = \frac{3\pi}{8} - \frac{1}{2},\quad \braket{r^2} = \frac{1}{2}; \\
    &\braket{\cos \theta} = - \frac{9\pi}{64}, \quad \braket{\cos 2\theta} = 0.
\end{align}
While the factor $\left| \eta \right|^2$ in Eq.~\eqref{eq:p-N3} reflects the level repulsion between $z_1$ and $z_2$,
the factor $\left| 1-\eta \right|^2$ encodes that between $z_2$ and $z_3$.
Additionally, $\left| 1-\eta \right|^2 = 1 + r^2 - 2r\cos \theta$ also captures the angular correlations:
configurations in which $z_2$ and $z_3$ are aligned in nearly the same direction $\theta \simeq 0$ are suppressed,
whereas those in which they lie in nearly the opposite directions $\theta \simeq \pi$ are favored.

\subsection{\texorpdfstring{$N=4$}{N=4}}

For $N=4$, the Vandermonde determinant factorizes as
\begin{align}
    &\Delta_4 \left( 0, z_2, z_3, z_4 \right) \nonumber \\
    &\quad= -\Delta_3 \left( 0, z_2, z_3 \right) \times z_4 \left( z_4 - z_2 \right) \left( z_4 - z_3 \right).
        \label{eq:Vandermonde_N4}
\end{align}
The integral over $z_4$ in Eq.~\eqref{eq:pN-t} then becomes
\begin{align}
    &\int_{\left| z_4\right|^2 > t} d^2 z_4~e^{-\left| z_4 \right|^2} \left| z_4 \right|^2 | z_4 - \eta \sqrt{t} |^2\,| z_4 - \sqrt{t} |^2 \nonumber \\
    &= \int_{t}^{\infty} \frac{ds}{2} e^{-s} s \int_{-\pi}^{\pi} d\phi \left| \sqrt{s} e^{\ii \phi} - \eta \sqrt{t} \right|^2 \left| \sqrt{s} e^{\ii \phi} - \sqrt{t} \right|^2 \nonumber \\
    &= \pi \int_{t}^{\infty} ds\,e^{-s} \left[ s^3 + t \left( 2+2u-v\right) s^2 + u t^2 s\right] \nonumber \\
    &= \pi e^{-t} \left[ 6 + \left( 10+4u-2v \right) t + \left( 7+5u-2v \right) t^2 \right. \nonumber \\
    &\qquad\qquad\qquad\qquad\qquad \left.+ \left( 3+3u-v \right) t^3 \right]
        \label{eq:spectator_N4}
\end{align}
with 
\begin{equation}
    u \coloneqq \left| \eta \right|^2 = r^2, \quad v \coloneqq \left| 1-\eta \right|^2 = 1+r^2-2r\cos\theta.
\end{equation}
Substituting this result into Eq.~\eqref{eq:pN-t},
we obtain
\begin{align}
    &p_4^{(0)} \left( \eta \right) \propto uv\int_0^\infty
    dt\,e^{-(2+u)\,t} \left[ 6t^4 + \left( 10+4u-2v \right) t^5 \right. \nonumber \\
    &\qquad\qquad \left.+ \left( 7+5u-2v \right) t^6 + \left( 3+3u-v \right) t^7 \right] \nonumber \\
    &=uv \left[ \frac{144}{\left(2+u \right)^5} + \frac{240 \left( 5+2u-v\right)}{\left(2+u \right)^6} \right. \nonumber \\
    &\qquad\left. + \frac{720 \left( 7+5u-2v \right)}{\left( 2+u\right)^7} + \frac{5040 \left( 3+3u-v \right)}{\left( 2+u\right)^8} \right].
        \label{eq:int_N4}
\end{align}
After imposing the normalization condition,
we arrive at
\begin{align}
    &p_4^{(0)} \left( \eta \right) = \frac{4uv}{\pi \left( 2+u \right)^8} \left( 13u^3-5u^2v+158u^2 \right. \nonumber \\
    &\qquad\qquad\qquad\quad \left. -50uv+746u-185v+649\right).
        \label{eq:p-N4}
\end{align}
As in the $N=3$ case, 
the level-repulsion factor $uv$ remains.
The additional polynomial factor arises from the presence of the remaining eigenvalue $z_4$,
which is constrained to lie outside the disk centered at the origin $z_1 = 0$ with the radius $\left| z_3 \right|$.

\subsection{Arbitrary \texorpdfstring{$N$}{N}}

For arbitrary $N\geq 3$, the factorization in Eq.~\eqref{eq:Vandermonde_N4} generalizes to
\begin{align}
&\left|\Delta_N(0,\eta\sqrt{t},\sqrt{t},z_4,\cdots,z_N)\right|^2
=uvt^3\left|\Delta_{N-3}(z_4,\cdots,z_N)\right|^2
\nonumber  \\
&\qquad\qquad\qquad\times\prod_{n=4}^N
|z_n|^2|z_n-\eta\sqrt{t}|^2|z_n-\sqrt{t}|^2.
\label{eq:class-A-general-vandermonde}
\end{align}
Although each remaining eigenvalue $z_n$ ($n \geq 4$) has the same weight as in Eq.~\eqref{eq:spectator_N4}, 
the factor $|\Delta_{N-3}|^2$ additionally couples these eigenvalues to each other. 
Expressing this factor as a product of two determinants and applying Andr\'eief's integral identity~\cite{Akemann-26}, 
we obtain
\begin{align}
&p_N^{(0)}(\eta)\propto{}uv\int_0^\infty dt\,t^4e^{-(1+u)t}
\nonumber \\
&\qquad\times\det_{1\leq i,j\leq N-3}\left[
\begin{aligned}
&\int_{|z|^2>t}d^2z\,e^{-|z|^2}z^i\bar z^j
\\
&\qquad\times|z-\eta\sqrt{t}|^2|z-\sqrt{t}|^2
\end{aligned}
\right].
\end{align}
For $N=3$, the empty determinant is defined as unity; 
for $N=4$, its single entry is precisely Eq.~\eqref{eq:spectator_N4}.

The matrix elements follow from the Gaussian integral
\begin{equation}
\int_{|z|^2>t}d^2z\,e^{-|z|^2}z^i\bar z^j
=
\pi e^{-t} \delta_{ij} i!\sum_{k=0}^{i}\cfrac{t^k}{k!} 
    \label{eq:class-A-exterior-moments}
\end{equation}
for nonnegative integers $i,j$. 
Since $|z-\eta\sqrt{t}|^2|z-\sqrt{t}|^2$ contains angular harmonics only up to second order, 
matrix elements with $|i-j|>2$ vanish.
Thus, the remaining integrals reduce to a pentadiagonal determinant, 
leaving only the integral over $t$.

An equivalent approach to obtaining an explicit algebraic expression is to expand the original Vandermonde determinant before applying Andr\'eief's identity.
After setting $z_1 = 0$, 
the row associated with $z_1$ selects the constant monomial,
so that the remaining monomial powers are $1,\ldots,N-1$.
Expanding the determinant along the rows associated with $z_2$ and $z_3$, 
and selecting two powers $1\leq i<j\leq N-1$, 
we obtain
\begin{equation}
(\eta\sqrt{t})^i(\sqrt{t})^j
-(\eta\sqrt{t})^j(\sqrt{t})^i
=t^{(i+j)/2}(\eta^i-\eta^j).
    \label{eq:32}
\end{equation}
For a fixed pair $(i, j)$, 
the remaining minor contains the complementary monomial powers in $\{1,\ldots,N-1\}$.
Upon taking the absolute square and integrating over the remaining eigenvalues $z_n$ ($n \geq 4$),
Eq.~\eqref{eq:class-A-exterior-moments} makes monomials of unequal powers orthogonal.
Consequently, cross terms between different pairs $(i, j)$ vanish exactly,
and each surviving diagonal term gives 
\begin{equation}
(N-3)!
\prod_{\substack{1\leq n\leq N-1\\ n\neq i,j}}
\left[
\pi e^{-t}n!
\sum_{l=0}^{n}\frac{t^l}{l!}
\right].
\end{equation}
The factor $e^{-(N-3)t}$ from these remaining integrals combines with the Gaussian factor $e^{-(1+u)t}$ of $z_2$ and $z_3$, giving $e^{-(N-2+u)t}$. 
Similarly, the factor $t^{i+j}$ from Eq.~\eqref{eq:32},
as well as the Jacobian factor $t$ in Eq.~\eqref{eq:pN-t}, gives $t^{i+j+1}$.

\begin{figure}[t]
\centering
\includegraphics[width=1.0\linewidth]{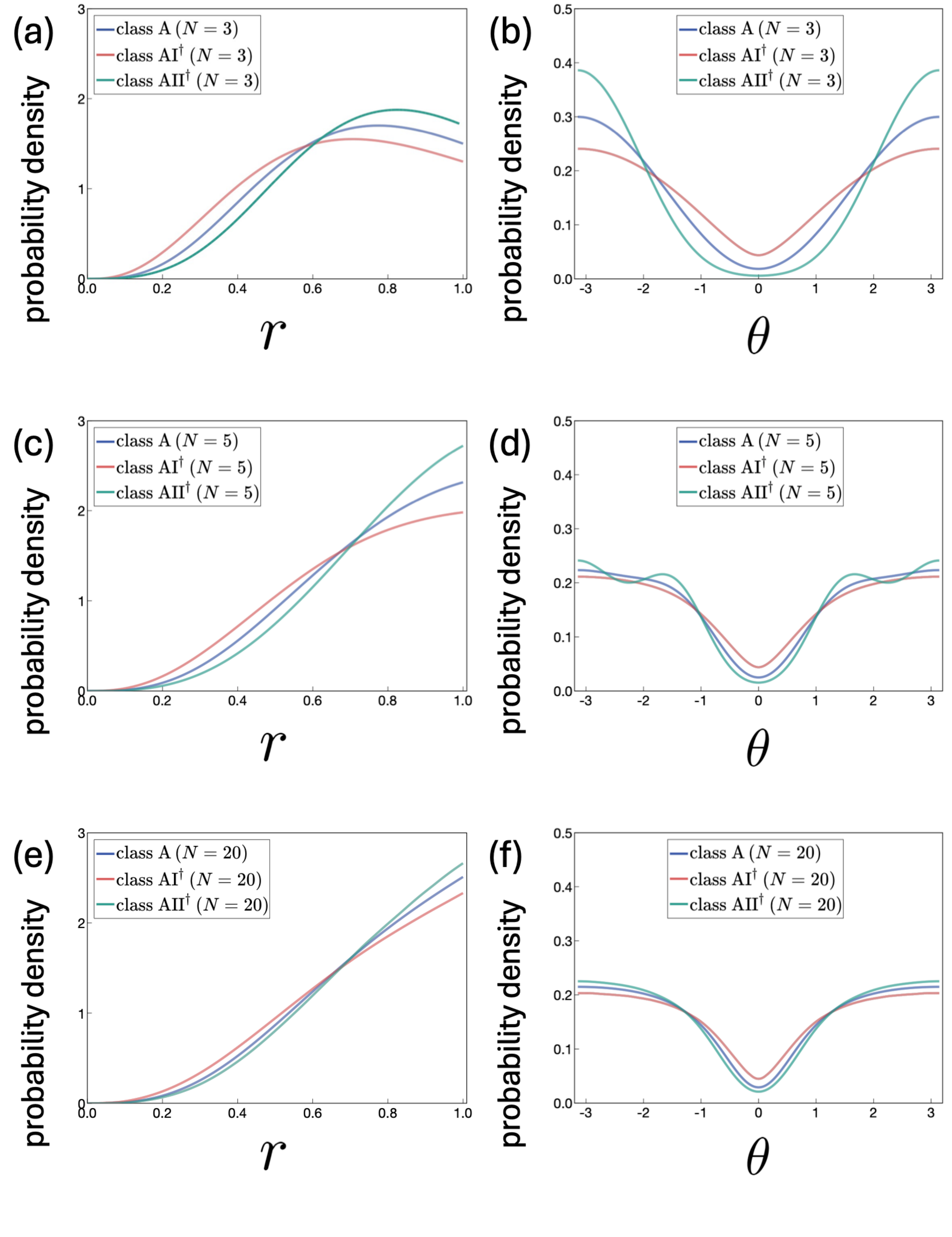} 
\caption{Probability densities of the modulus $r$ and argument $\theta$ of complex spacing ratios $re^{\ii \theta}$ for the Gaussian ensembles of non-Hermitian random matrices in classes A, AI$^{\dag}$, and AII$^{\dag}$ with (a, b)~$N=3$, (c, d)~$N=5$, and (e, f)~$N=20$.
Class A: analytical results in Eq.~\eqref{eq:class-A-general-algebraic} (blue).
Class AI$^{\dag}$: analytical results in Eq.~\eqref{eq:p_exact_A1_N3} for $N=3$ and Monte Carlo estimates based on Eq.~\eqref{eq:MCMC} for $N=5, 20$ (red).
Class AII$^{\dag}$: analytical results in Eq.~\eqref{eq:class-A2-general-algebraic} for $N=3, 5$ and numerical results for $N=20$ (green).}	
    \label{fig:RMT}
\end{figure}

The integral of the unnormalized origin-conditioned density is
\begin{align}
&\int
d^2z_2 \cdots d^2z_N\,
e^{-\sum_{n=2}^{N}|z_n|^2}
\left|\Delta_N(0,z_2,\ldots,z_N)\right|^2
\nonumber\\
&\qquad\qquad\qquad\qquad=
(N-1)!\pi^{N-1}
\prod_{n=1}^{N-1}n!.
\end{align}
Including the $(N-1)(N-2)$ possible assignments of the nearest and next-to-nearest neighbors, 
we obtain the normalized expression
\begin{align}
&p_N^{(0)}(\eta)
=\frac{1}{\pi}\sum_{1\leq i<j\leq N-1}
\frac{|\eta^i-\eta^j|^2}{i!\,j!}
\nonumber \\
&\quad\times\int_0^\infty dt\,t^{i+j+1}e^{-(N-2+u)t} \prod_{\substack{1\leq n\leq N-1\\n\ne i,j}}
\left(\sum_{l=0}^{n}\frac{t^l}{l!}\right).
\label{eq:class-A-general-pair-integral}
\end{align}
The product is a polynomial of degree $N(N-1)/2-i-j$. 
Expanding it in $t$ and integrating each term as in Eq.~\eqref{eq:int_N4} gives
\begin{align}
&p_N^{(0)}(\eta)
=\frac{1}{\pi}\sum_{1\leq i<j\leq N-1}
\frac{|\eta^i-\eta^j|^2}{i!\,j!}
\nonumber \\
&\qquad\times\sum_{k=0}^{N(N-1)/2-i-j}
\frac{(i+j+k+1)!}{k!\,(N-2+u)^{i+j+k+2}}
\nonumber \\
&\qquad\times\left.\frac{d^k}{dt^k}
\prod_{\substack{1\leq n\leq N-1\\n\ne i,j}}
\left(\sum_{l=0}^{n}\frac{t^l}{l!}\right)
\right|_{t=0}.
    \label{eq:class-A-general-algebraic}
\end{align}
These expressions are exact for the origin-conditioned distribution at each finite $N$.
The derivative at $t=0$, 
divided by $k!$, 
extracts the corresponding polynomial coefficient. 
Each term in Eq.~\eqref{eq:class-A-general-algebraic} contains the same level-repulsion factor
$uv=|\eta|^2|1-\eta|^2$, 
since $|\eta^i-\eta^j|^2$ is divisible by $uv$ for $1\leq i<j\leq N-1$.
The single pair $(i,j)=(1,2)$ for $N=3$ reproduces Eq.~\eqref{eq:p-N3}, 
while the three pairs for $N=4$ reproduce Eq.~\eqref{eq:p-N4}. 
In Fig.~\ref{fig:RMT}, we plot the radial and angular marginal densities obtained from Eq.~\eqref{eq:class-A-general-algebraic} for $N=3, 5, 20$.
In Table~\ref{tab:analytical}, 
we provide representative moments for several finite values of $N$,
evaluated from Eq.~\eqref{eq:class-A-general-algebraic}.

\begin{table*}[t]
	\centering
	\caption{Finite-$N$ moments of complex spacing ratios $re^{\ii\theta}$ for the Gaussian ensembles of non-Hermitian random matrices in classes A, AI$^{\dag}$, and AII$^{\dag}$.
    For class AI$^{\dag}$,
    the analytical results in Eq.~\eqref{eq:p_exact_A1_N3} are shown for $N=3$,
    whereas the Monte Carlo estimates based on Eq.~\eqref{eq:MCMC} are shown for $N=4, 5, 10, 20$,
    with the uncertainty representing the standard error.}
     \begin{tabular}{cc|cccc} \hline \hline
     ~~Class~~ & ~~$N$~~ & ~~$\braket{r}$~~ & ~~$\braket{r^2}$~~& ~~$-\braket{\cos \theta}$~~ & ~~$-\braket{\cos 2\theta}$~~  \\ \hline
     A~\cite{Dusa-22, Akemann-26} & $3$ & ~~$0.678097$~~ & ~~$0.5$~~ & ~~$0.441786$~~ & ~~$0$~~ \\
     & $4$ & $0.718234$ & $0.552726$ & $0.319756$ & $0.111111$ \\
     & $5$ & $0.731102$ & $0.570306$ & $0.275250$ & $0.109875$ \\
     & $10$ & $0.738620$ & $0.580792$ & $0.247024$ & $0.100269$ \\
     & $20$ & $0.738660$ & $0.580849$ & $0.246831$ & $0.100172$ \\ \hline
     AI$^{\dag}$ [Eqs.~\eqref{eq:p_exact_A1_N3} and \eqref{eq:MCMC}] & $3$ & ~~$0.645811$~~ & ~~$0.462513$~~ & ~~$0.289824$~~ & ~~$0.044689$~~ \\
     & $4$ & ~~$0.686674 \pm 0.000099$~~ & ~~$0.513883 \pm 0.000075$~~ & ~~$0.244949 \pm 0.000065$~~ & ~~$0.077743 \pm 0.000045$~~ \\
     & $5$ & $0.70180 \pm 0.00019$ & $0.53352 \pm 0.00023$ & $0.22245 \pm 0.00054$ & $0.08362 \pm 0.00044$ \\
     & $10$ & $0.71827 \pm 0.00030$ & $0.55536 \pm 0.00033$ & $0.19905 \pm 0.00050$ & $0.08495 \pm 0.00050$ \\
     & $20$ & $0.72111 \pm 0.00042$ & $0.55901 \pm 0.00051$ & $0.19668 \pm 0.00074$ & $0.08375 \pm 0.00060$ \\ \hline
     AII$^{\dag}$ [Eq.~\eqref{eq:class-A2-general-algebraic}] & $3$ & ~~$0.705482$~~ & ~~$0.533565$~~ & ~~$0.597230$~~ & ~~$-0.115741$~~ \\
     & $4$ & $0.746621$ & $0.589824$ & $0.366456$ & $0.170846$ \\
     & $5$ & $0.757387$ & $0.605470$ & $0.298304$ & $0.135222$ \\
     & $6$ & $0.758839$ & $0.607897$ & $0.278052$ & $0.115620$ \\ \hline \hline
    \end{tabular}
	\label{tab:analytical}
\end{table*}

\section{Class AII\texorpdfstring{$^{\dagger}$}{†}}
    \label{sec:AII}

For the Gaussian ensemble in class AII$^{\dag}$, 
the joint eigenvalue probability density function reads~\cite{Xiao-26}
\begin{align}
    &\rho_N \left( z_1, \cdots, z_N \right) \propto e^{- \sum_{n=1}^{N}
    \left| z_n \right|^2} \nonumber \\ 
    &\qquad\times \left| \Delta_N \left( z_1, \cdots, z_N \right) \right|^2 R_N \left( z_1, \cdots, z_N \right), 
        \label{eq:A2-jpdf}
\end{align}
with the Vandermonde determinant $\Delta_N$ in Eq.~\eqref{eq:Vandermonde}.
Whereas $\Delta_N$ consists solely of the products of the pairwise differences $z_i - z_j$,
the polynomial $R_N$ contains generic many-body contributions among $z_1$, $\cdots$, $z_N$ (see Ref.~\cite{Xiao-26} for details;
see also below for $N=3, 4$).
Here, $N$ denotes the number of different eigenvalues,
and $2N$ denotes the matrix dimension.
On the basis of this joint eigenvalue density, 
we derive the complex-spacing-ratio distributions for finite $N$,
first explicitly for $N=3, 4$ and subsequently for generic $N$.
In addition to these finite-$N$ analytical results, 
we numerically diagonalize non-Hermitian random matrices drawn from the Gaussian ensemble and obtain the complex-spacing-ratio distributions for several values of $N$,
as summarized in Fig.~\ref{fig:AII} and Table~\ref{tab:numerical} (see Appendix~\ref{appendix:origin} for details on numerical calculations of origin-conditioned complex-spacing-ratio distributions).
The analytically evaluated finite-$N$ moments listed in Table~\ref{tab:analytical} agree with those obtained from direct numerical diagonalization in Table~\ref{tab:numerical},
with all deviations within two combined standard errors.

\begin{figure}[t]
\centering
\includegraphics[width=1.0\linewidth]{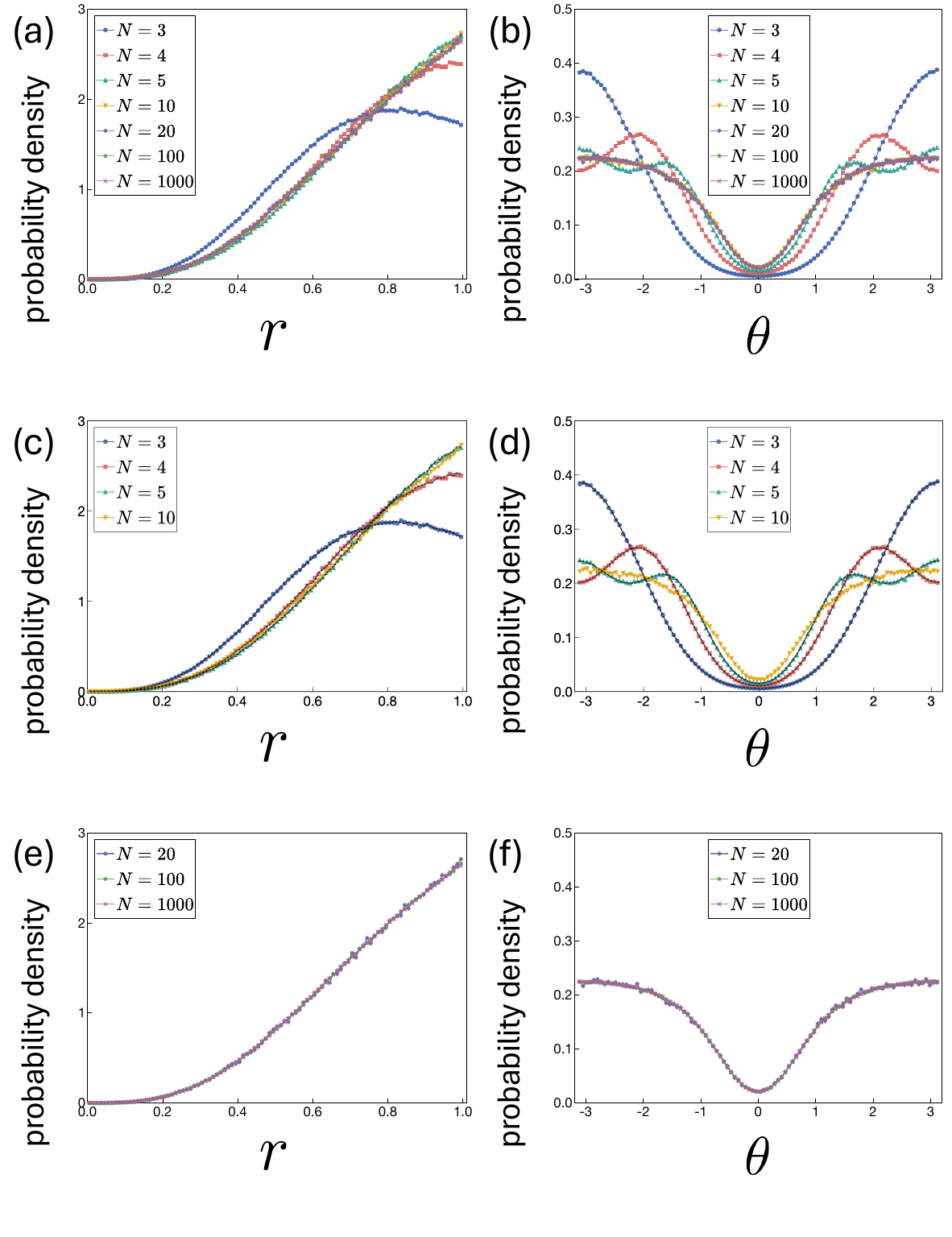} 
\caption{Probability densities of the modulus $r$ and argument $\theta$ of complex spacing ratios $re^{\ii \theta}$ for the Gaussian ensemble of non-Hermitian random matrices in class AII$^{\dag}$ with (a, b)~$N=3, 4, 5, 10, 20, 100, 1000$, (c, d)~$N=3, 4, 5, 10$, and (e, f)~$N=20, 100, 1000$.
The distributions are numerically obtained from $5 \times 10^6$ realizations for $N \leq 20$,
$10^5$ realizations for $N=100$, 
and $10^4$ realizations for $N=1000$.
Whereas the origin-conditioned distribution is considered for $N \leq 20$,
the central $50\%$ of the complex spectrum is employed for $N > 20$.
(c, d)~Black dashed curves: 
exact finite-$N$ expressions in Eq.~\eqref{eq:class-A2-general-algebraic} for $N=3,4,5$.}	
    \label{fig:AII}
\end{figure}

\begin{table*}[t]
	\centering
	\caption{Moments of complex spacing ratios $re^{\ii\theta}$ obtained by numerical diagonalization of the Gaussian ensembles of non-Hermitian random matrices in classes AI$^{\dag}$ and AII$^{\dag}$.
    For $N \leq 20$, 
    the moments are numerically evaluated from the origin-conditioned distribution using $5 \times 10^6$ realizations.
    For $N > 20$, they are evaluated using the central $50\%$ of the complex spectrum, 
    with $10^5$ realizations for $N=100$, $5 \times 10^4$ realizations for $N=200$, $2 \times 10^4$ realizations for $N=500$, and 
    $10^4$ realizations for $N=1000$.
    The uncertainty represents the standard error.}
     \begin{tabular}{cc|cccc} \hline \hline
     ~~Class~~ & ~~$N$~~ & ~~$\braket{r}$~~ & ~~$\braket{r^2}$~~& ~~$-\braket{\cos \theta}$~~ & ~~$-\braket{\cos 2\theta}$~~  \\ \hline
     AI$^{\dag}$ & $3$ & ~~$0.64575 \pm 0.00010$~~ & ~~$0.46245 \pm 0.00013$~~ & ~~$0.28962 \pm 0.00026$~~ & ~~$0.04526 \pm 0.00036$~~ \\
     & $4$ & ~~$0.68670 \pm 0.00012$~~ & ~~$0.51385 \pm 0.00015$~~ & ~~$0.24527 \pm 0.00031$~~ & ~~$0.07757 \pm 0.00037$~~ \\
     & $5$ & ~~$0.70197 \pm 0.00013$~~ & ~~$0.53381 \pm 0.00016$~~ & ~~$0.22300 \pm 0.00035$~~ & ~~$0.08361 \pm 0.00040$~~ \\
     & $10$ & ~~$0.71830 \pm 0.00016$~~ & ~~$0.55540 \pm 0.00021$~~ & ~~$0.19911 \pm 0.00047$~~ & ~~$0.08445 \pm 0.00053$~~ \\
     & $20$ & ~~$0.72141 \pm 0.00022$~~ & ~~$0.55951 \pm 0.00028$~~ & ~~$0.19578 \pm 0.00066$~~ & ~~$0.08451 \pm 0.00072$~~ \\
     & $100$ & ~~$0.72226 \pm 0.00011$~~ & ~~$0.56060 \pm 0.00014$~~ & ~~$0.19526 \pm 0.00033$~~ & ~~$0.08470 \pm 0.00033$~~ \\
     & $200$ & ~~$0.72214 \pm 0.00011$~~ & ~~$0.56044 \pm 0.00014$~~ & ~~$0.19518 \pm 0.00033$~~ & ~~$0.08478 \pm 0.00033$~~ \\
     & $500$ & ~~$0.72227 \pm 0.00011$~~ & ~~$0.56066 \pm 0.00014$~~ & ~~$0.19541 \pm 0.00033$~~ & ~~$0.08416 \pm 0.00033$~~ \\
     & $1000$ & ~~$0.72236 \pm 0.00011$~~ & ~~$0.56071 \pm 0.00014$~~ & ~~$0.19484 \pm 0.00033$~~ & ~~$0.08370 \pm 0.00033$~~ \\ \hline
     AII$^{\dag}$ & $3$ & ~~$0.70551 \pm 0.00013$~~ & ~~$0.53356 \pm 0.00018$~~ & ~~$0.59736 \pm 0.00027$~~ & ~~$-0.11558 \pm 0.00048$~~ \\
     & $4$ & ~~$0.74642 \pm 0.00016$~~ & ~~$0.58952 \pm 0.00022$~~ & ~~$0.36706 \pm 0.00042$~~ & ~~$0.17099 \pm 0.00054$~~ \\
     & $5$ & ~~$0.75707 \pm 0.00018$~~ & ~~$0.60509 \pm 0.00024$~~ & ~~$0.29834 \pm 0.00057$~~ & ~~$0.13485 \pm 0.00069$~~ \\
     & $6$ & ~~$0.75869 \pm 0.00020$~~ & ~~$0.60772 \pm 0.00027$~~ & ~~$0.27794 \pm 0.00064$~~ & ~~$0.11592 \pm 0.00071$~~ \\
     & $10$ & ~~$0.75236 \pm 0.00026$~~ & ~~$0.59903 \pm 0.00035$~~ & ~~$0.27995 \pm 0.00084$~~ & ~~$0.10508 \pm 0.00096$~~ \\
     & $20$ & ~~$0.74973 \pm 0.00036$~~ & ~~$0.59555 \pm 0.00050$~~ & ~~$0.2846 \pm 0.0012$~~ & ~~$0.1068 \pm 0.0014$~~ \\
     & $100$ & ~~$0.74913 \pm 0.00010$~~ & ~~$0.59458 \pm 0.00013$~~ & ~~$0.28515 \pm 0.00031$~~ & ~~$0.10671 \pm 0.00032$~~ \\
     & $200$ & ~~$0.74899 \pm 0.00010$~~ & ~~$0.59441 \pm 0.00014$~~ & ~~$0.28591 \pm 0.00031$~~ & ~~$0.10571 \pm 0.00032$~~ \\
     & $500$ & ~~$0.74905 \pm 0.00010$~~ & ~~$0.59448 \pm 0.00013$~~ & ~~$0.28686 \pm 0.00031$~~ & ~~$0.10565 \pm 0.00032$~~ \\
     & $1000$ & ~~$0.74892 \pm 0.00010$~~ & ~~$0.59433 \pm 0.00014$~~ & ~~$0.28573 \pm 0.00032$~~ & ~~$0.10695 \pm 0.00032$~~ \\ \hline \hline
    \end{tabular}
	\label{tab:numerical}
\end{table*}

\subsection{\texorpdfstring{$N=3$}{N=3}}

For $N=3$, the polynomial $R_3$ in Eq.~\eqref{eq:A2-jpdf} is given as~\cite{Xiao-26}
\begin{equation}
    R_3 \left( z_1, z_2, z_3 \right) = \prod_{1 \leq i < j \leq 3} \left( 1 + \frac{\left| z_i - z_j\right|^2}{2}\right) + \frac{1}{2}.
\end{equation}
In the absence of the last term $1/2$, 
$R_3$ would be constructed entirely from factors depending only on the pairwise eigenvalue separations $\left| z_i - z_j \right|$,
in a similar manner to the Vandermonde determinant.
The presence of the additional term $1/2$ gives rise to the nontrivial eigenvalue correlations that cannot be reduced solely to two-body interactions.
Substituting $z_1 = 0$, $z_2 = \eta \sqrt{t}$, and $z_3 = \sqrt{t}$ into $R_3$, 
we have
\begin{align}
    R_3 &= \left( 1 + \frac{t}{2}\right) \left( 1 + \frac{ut}{2}\right) \left( 1 + \frac{vt}{2}\right) + \frac{1}{2} \nonumber \\
    &= \frac{3}{2} + \frac{1+u+v}{2} t + \frac{u+v+uv}{4} t^2 + \frac{uv}{8} t^3,
\end{align}
leading to
\begin{align}
    &\rho_3^{(0)}\,( \eta \sqrt{t}, \sqrt{t} ) \propto uv t^3 e^{-(1+u)\,t} \nonumber \\
    &\quad\times \left( \frac{3}{2} + \frac{1+u+v}{2} t + \frac{u+v+uv}{4} t^2 + \frac{uv}{8} t^3 \right).
\end{align}

Then, Eq.~\eqref{eq:pN-t} yields
\begin{align}
    &p_3^{(0)} \left( \eta \right) \propto uv \int_0^{\infty} dt\,t^4 e^{-(1+u)\,t} \nonumber \\
    &\quad \times \left( \frac{3}{2} + \frac{1+u+v}{2} t + \frac{u+v+uv}{4} t^2 + \frac{uv}{8} t^3 \right) \nonumber \\
    &= uv \left[ \frac{36}{\left( 1+u \right)^5} + \frac{60 \left( 1+u+v \right)}{\left( 1+u \right)^6} \right. \nonumber \\
    &\qquad\qquad\qquad\left.+ \frac{180 \left( u+v+uv \right)}{\left( 1+u \right)^7} + \frac{630uv}{\left( 1+u \right)^8} \right].
\end{align}
Imposing the normalization condition, we obtain
\begin{align}
    &p_3^{(0)} \left( \eta \right) = \frac{uv}{9\pi \left( 1+u \right)^8} \left( 16u^3 + 40u^2v +78u^2 \right. \nonumber \\
    &\qquad\qquad\qquad\left. + 185uv +78u + 40v + 16\right).
        \label{eq:p_exact_N3}
\end{align}
Compare this result with the counterpart for class A in Eq.~\eqref{eq:p-N3}.
While the level-repulsion factor $uv = \left| \eta \right|^2 \left| 1-\eta \right|^2$ is common to both classes,
the remaining factor is changed by time-reversal symmetry$^{\dag}$.
The radial and angular marginal densities are obtained as
\begin{align}
    p_{r, 3}^{(0)} \left( r \right) &= \frac{2r^3 \left( 56+439r^2 + 976r^4 + 439r^6 +56r^8 \right)}{9 \left( 1+r^2 \right)^8}, \label{eq:p_r_N3_A2} \\
    p_{\theta, 3}^{(0)} \left( \theta \right) &= \frac{1}{2\pi} \left( 1 - \frac{73\pi}{192} \cos \theta + \frac{25}{108} \cos 2 \theta \right),
        \label{eq:p_theta_N3_A2}
\end{align}
with the representative moments
\begin{align}
    &\braket{r} = \frac{6689\pi}{18432} - \frac{751}{1728},\quad \braket{r^2} = \frac{461}{864}; \\
    &\braket{\cos \theta} = - \frac{73\pi}{384}, \quad \braket{\cos 2\theta} = \frac{25}{216}.
\end{align}
In Fig.~\ref{fig:RMT}\,(a, b), we plot $p_{r, 3}^{(0)} \left( r \right)$ and $p_{\theta, 3}^{(0)} \left( \theta \right)$ for class AII$^{\dag}$,
in addition to the counterparts for classes A and AI$^{\dag}$.
Despite the common cubic behavior $p_{r, 3} \left( r \right) \propto r^3$ ($0 < r \ll 1$) between classes A and AII$^{\dag}$,
the mean ratio $\braket{r}$ is larger in class AII$^{\dag}$ than in class A,
implying the enhanced level repulsion induced by time-reversal symmetry$^{\dag}$.
This is also consistent with more negative $\braket{\cos \theta}$ in class AII$^{\dag}$ relative to class A.
It is also notable that the second angular harmonic $\cos 2\theta$ is generated in $p_{\theta, 3}$;
compare Eq.~\eqref{eq:p_theta_N3_A2} with Eq.~\eqref{eq:p_theta_N3_A}.
Among the cases investigated in Tables~\ref{tab:analytical} and \ref{tab:numerical},
$\braket{\cos 2\theta} > 0$ occurs only for $N=3$ in class AII$^{\dag}$.

\subsection{\texorpdfstring{$N=4$}{N=4}}

For $N=4$, the polynomial $R_4$ in Eq.~\eqref{eq:A2-jpdf} is given as~\cite{Xiao-26}
\begin{align}
    &R_4 \left( z_1, z_2, z_3, z_4 \right) = \prod_{1\leq i < j \leq 4} \left( 1+ \frac{\left| z_i - z_j \right|^2}{2} \right) \nonumber \\
    &\qquad + \frac{1}{2} \sum_{i=1}^{4} \prod_{1\leq j \leq 4, j \neq i} \left( 1+ \frac{\left| z_i - z_j \right|^2}{2} \right) \nonumber \\
    &\qquad+ \frac{1}{4} \sum_{1\leq i < j \leq 4} \left( 1+ \frac{\left| z_i - z_j \right|^2}{2} \right).
        \label{eq:R4}
\end{align}
As in the $N=3$ case,
$R_4$ does not factorize into products depending solely on the pairwise separations $\left| z_i - z_j \right|$.
Using the Vandermonde determinant
\begin{align}
    &\left| \Delta_4\,( 0, \eta \sqrt{t}, \sqrt{t}, z_4 ) \right|^2 \nonumber \\
    &\quad= uvt^3 \left| z_4 \right|^2 | z_4 - \sqrt{t} |^2\,| z_4 - \eta \sqrt{t} |^2,
\end{align}
we obtain the complex-spacing-ratio distribution in Eq.~\eqref{eq:pN-t} as
\begin{align}
    &p_4^{(0)} \left( \eta \right) \propto uv \int_0^{\infty} dt\,t^4 e^{-(1+u)\,t} \nonumber \\
    &\times \int_{\left| z_4 \right|^2 > t} d^2 z_4\,e^{-|z_4|^2} \left| z_4 \right|^2 | z_4 - \sqrt{t} |^2\,| z_4 - \eta \sqrt{t} |^2 \nonumber \\
    &\qquad\qquad\qquad\qquad\times R_4 \,( 0, \eta \sqrt{t}, \sqrt{t}, z_4 ) .
\end{align}

We next expand the polynomial in the integrand by
\begin{align}
    &\left| z_4 \right|^2 | z_4 - \sqrt{t} |^2\,| z_4 - \eta \sqrt{t} |^2 R_4\,( 0, \eta \sqrt{t}, \sqrt{t}, z_4 ) \nonumber \\
    &\quad\eqqcolon \sum_{i, j = 1}^{6} c_{ij} \left( t, \eta, \eta^* \right) z_4^i \left( z_4^* \right)^j.
        \label{eq:c4}
\end{align}
From Eq.~\eqref{eq:class-A-exterior-moments}, the integral over $z_4$ can be written as
\begin{align}
    &\int_{\left| z_4 \right|^2 > t} d^2 z_4\,e^{-|z_4|^2} \left| z_4 \right|^2 | z_4 - \sqrt{t} |^2\,| z_4 - \eta \sqrt{t} |^2 \nonumber \\
    &\qquad\qquad\qquad\qquad \times R_4 \,( 0, \eta \sqrt{t}, \sqrt{t}, z_4 )  \nonumber \\
    &\qquad = \pi e^{-t} \sum_{i=1}^{6} c_{ii} \left( t, \eta, \eta^{*} \right) i! \sum_{k=0}^{i} \frac{t^k}{k!}.
\end{align}
Hereafter, using $\eta \eta^{*} = u$ and $\eta + \eta^{*} = 1+u-v$,
we express $c_{ii} \left( t, \eta, \eta^{*} \right)$ as a polynomial of $t$, $u$, and $v$,
denoted by $c_{ii} \left( t, u, v \right)$.
For example, we have
\begin{equation}
    c_{66} = \frac{1}{8} \left[ \left( 1+ \frac{t}{2} \right) \left( 1+ \frac{ut}{2} \right) \left( 1+ \frac{vt}{2} \right) + \frac{1}{2} \right].
\end{equation}

Further introducing the polynomial $a_{l} \left( u, v \right)$ through
\begin{equation}
    \sum_{l=0}^{9} a_{l} \left( u, v \right) t^l \coloneqq \sum_{i=1}^6 c_{ii} \left( t, u, v \right) i! \sum_{k=0}^{i} \frac{t^k}{k!},
        \label{eq:a4}
\end{equation}
we finally obtain
\begin{align}
    p_4^{(0)} \left( \eta \right) &\propto uv \sum_{l=0}^{9} a_l \left( u, v \right) \int_0^{\infty} dt\,t^{l+4} e^{-(2+u)\,t} \nonumber\\
    &= uv \sum_{l=0}^{9} \frac{\left( l+4 \right)!\,a_l \left( u, v \right)}{\left( 2+u \right)^{l+5}}.
\end{align}
Imposing the normalization condition yields
\begin{equation}
    p_4^{(0)} \left( \eta \right) = \frac{uv}{32400\pi} \sum_{l=0}^{9} \frac{\left( l+4 \right)!\,a_l \left( u, v \right)}{\left( 2+u \right)^{l+5}}.
        \label{eq:p_exact_N4}
\end{equation}
Notably, the level-repulsion factor $uv$ remains explicit,
as in class A.
The polynomials $a_l \left( u, v \right)$ can be obtained straightforwardly by symbolic computation from Eqs.~\eqref{eq:R4}, \eqref{eq:c4}, and \eqref{eq:a4},
and are summarized in Table~\ref{tab:AII-N4-a}.
In this manner, the moments of $r$ and $\theta$ are also evaluated analytically, as in
\begin{align}
    &\braket{r} = -\frac{300249322238339}{352241005363200} \nonumber \\
    &\qquad\qquad+\frac{591760511489\sqrt{2}}{322122547200} \arctan\frac{1}{\sqrt{2}}, \\
    &\braket{r^2} = \frac{445805233}{755827200}; \\
    &\braket{\cos \theta} = \frac{99960759981257}{2377626786201600} \nonumber \\
    &\qquad\qquad -\frac{113381873003\sqrt{2}}{241591910400}
    \arctan\frac{1}{\sqrt{2}}, \\
    &\braket{\cos 2\theta} = -\frac{2324334173}{13604889600},
\end{align}
which agree with the numerical results summarized in Table~\ref{tab:numerical}.

\begin{table*}[t]
\caption{Polynomials $a_l(u,v)$ appearing in the exact $N=4$ complex-spacing-ratio distribution for class AII$^\dagger$ [Eq.~\eqref{eq:p_exact_N4}; $u\coloneqq|\eta|^2$ and $v\coloneqq|1-\eta|^2$].}
    \label{tab:AII-N4-a}
\centering
\small
\begin{tabular}{c l}
\hline
\hline
~~$l$~~ & $a_l(u,v)$ \\
\hline
$0$
&
$\displaystyle
378
$
\\[1mm]

$1$
&
$\displaystyle
42\left(22+13u-2v\right)
$
\\[1mm]

$2$
&
$\displaystyle
3\left(
64u^2+27uv+366u-20v^2-v+309
\right)
$
\\[1mm]

$3$
&
$\displaystyle
\frac{3}{4}\left(
24u^3+92u^2v+486u^2
-62uv^2+325uv+1330u
+3v^3-142v^2+144v+728
\right)
$
\\[2mm]

$4$
&
$\displaystyle
\frac{1}{8}\left(
72u^3v+258u^3
-66u^2v^2+1128u^2v+2524u^2
+9uv^3-636uv^2+2202uv+4388u
+27v^3-678v^2+848v+1733
\right)
$
\\[2mm]

$5$
&
$\displaystyle
\frac{1}{16}\left(
258u^3v+420u^3
-216u^2v^2+2056u^2v+2644u^2
+27uv^3-984uv^2+2781uv+3246u
+36v^3-658v^2+886v+960
\right)
$
\\[2mm]

$6$
&
$\displaystyle
\frac{1}{16}\left(
210u^3v+203u^3
-160u^2v^2+1124u^2v+912u^2
+18uv^3-466uv^2+1154uv+810u
+15v^3-217v^2+307v+173
\right)
$
\\[2mm]

$7$
&
$\displaystyle
\frac{1}{32}\left(
203u^3v+118u^3
-144u^2v^2+814u^2v+394u^2
+15uv^3-294uv^2+651uv+238u
+8v^3-94v^2+138v+30
\right)
$
\\[2mm]

$8$
&
$\displaystyle
\frac{1}{32}\left(
59u^3v+15u^3
-40u^2v^2+187u^2v+40u^2
+4uv^3-57uv^2+109uv+15u
+v^3-10v^2+15v
\right)
$
\\[2mm]

$9$
&
$\displaystyle
\frac{uv}{64}\left(
15u^2-10uv+40u+v^2-10v+15
\right)
$
\\
\hline
\hline
\end{tabular}
\end{table*}

\subsection{Arbitrary \texorpdfstring{$N$}{N}}

For generic $N$, we derive the complex-spacing-ratio distribution $p_N^{(0)} \left( \eta \right)$ in a similar manner to $N=4$.
For arbitrary $N$, $R_N$ in the joint eigenvalue density in Eq.~\eqref{eq:A2-jpdf} is a polynomial consisting of $\left| z_i - z_j \right|^2/2$.
Then, we focus on the polynomial part of the integrand in Eq.~\eqref{eq:pN-t}:
\begin{align}
    &uvt^3 \mathcal{C}_N \coloneqq \left| \Delta_N\,( 0, \eta \sqrt{t}, \sqrt{t}, z_4, \cdots, z_N ) \right|^2 \nonumber \\
    &\qquad\qquad\qquad \times R_N\,( 0, \eta\sqrt{t}, \sqrt{t}, z_4, \cdots, z_N ).
\end{align}
We expand $\mathcal{C}_N$ in powers of $z_4, \cdots, z_N$ and $z_4^*, \cdots, z_N^*$ by 
\begin{equation}
    \mathcal{C}_N = \sum_{{\bm m}, {\bm n}} c_{{\bm m}, {\bm n}}^{(N)} \left( t, \eta, \eta^* \right) \prod_{i=4}^{N} z_i^{m_i} \left( z_i^*\right)^{n_i}
        \label{eq:cN}
\end{equation}
with ${\bm m} \coloneqq \left( m_4, \cdots, m_N \right)$ and ${\bm n} \coloneqq \left( n_4, \cdots, n_N \right)$.
Using the Gaussian integral in Eq.~\eqref{eq:class-A-exterior-moments},
we have
\begin{align}
    &\int_{\left| z_i\right|^2 > t} \prod_{i=4}^N \left( d^2 z_i\,e^{-|z_i|^2}\right) \mathcal{C}_N \nonumber \\
    &= \pi^{N-3} e^{-(N-3)\,t} \sum_{\bm m} c_{{\bm m}, {\bm m}}^{(N)} \left( t, \eta, \eta^* \right) \prod_{i=4}^{N} \left( m_i! \sum_{k=0}^{m_i} \frac{t^k}{k!}\right),
\end{align}
and further introduce polynomials $a_{N, l} \left( u, v \right)$ through
\begin{align}
    &\sum_{\bm m} c_{{\bm m}, {\bm m}}^{(N)} \left( t, \eta, \eta^* \right) \prod_{i=4}^{N} \left( m_i! \sum_{k=0}^{m_i} \frac{t^k}{k!}\right) \nonumber \\
    &\qquad \eqqcolon \sum_{l=0}^{N(N-1) - 3} a_{N, l} \left( u, v \right) t^{l}.
        \label{eq:aN}
\end{align}
Here, the squared Vandermonde determinant $|\Delta_N|^2$ has degree $N \left( N-1 \right)/2$ in the squared pairwise separations $\left| z_i - z_j \right|^2$,
and $R_N$ has degree at most $N \left( N-1 \right)/2$.
After extracting the factor $uvt^3$,
the resulting polynomial has degree at most 
\begin{equation}
    2 \times \frac{N \left( N-1 \right)}{2} - 3 = N \left( N-1 \right) - 3,
\end{equation}
which determines the upper limit of $l$ in Eq.~\eqref{eq:aN}.

Equation~\eqref{eq:pN-t} then gives
\begin{align}
    p_N^{(0)} \left( \eta \right) 
    &\propto uv \sum_{l=0}^{N(N-1)-3} a_{N, l} \left( u, v \right) \int_0^{\infty}dt\,t^{l+4}e^{-(N-2+u)\,t} \nonumber \\
    &= uv \sum_{l=0}^{N(N-1)-3} a_{N, l} \left( u, v \right) \frac{\left( l+4 \right)!}{\left( N-2+u \right)^{l+5}}.
\end{align}
Imposing the normalization condition, we obtain
\begin{align}
    &p_N^{(0)} \left( \eta \right) = \frac{\left( N-1 \right) \left( N-2 \right) uv}{\pi D_N} \nonumber \\
    &\qquad\qquad\quad \times \sum_{l=0}^{N(N-1)-3} \frac{\left( l+4 \right)!\,a_{N, l} \left( u, v \right)}{\left( N-2+u \right)^{l+5}}
        \label{eq:class-A2-general-algebraic}
\end{align}
with 
\begin{align}
    &D_N \coloneqq \int \left( \prod_{n=2}^{N} \frac{d^2 z_n}{\pi} e^{-|z_n|^2}\right) \left| \Delta_N \left( 0, z_2, \cdots, z_N \right) \right|^2 \nonumber \\
    &\qquad\qquad\qquad\qquad\qquad \times R_N \left( 0, z_2, \cdots, z_N \right). 
\end{align}
For example, we have $D_3 = 108$, $D_4 = 194400$, and $D_5 = 17146080000$.

Equation~\eqref{eq:class-A2-general-algebraic} provides an algebraic expression of the complex-spacing-ratio distribution for arbitrary finite $N$,
consistent with Eq.~\eqref{eq:p_exact_N3} for $N=3$ and Eq.~\eqref{eq:p_exact_N4} for $N=4$.
As in the $N=4$ case,
we can obtain the polynomials $a_{N, l} \left( u, v \right)$ systematically through Eqs.~\eqref{eq:cN} and \eqref{eq:aN}.
Using this construction, we obtain the marginal densities $p_{r, N} \left( r \right)$ and $p_{\theta, N} \left( \theta \right)$ for $N=4, 5, 6$,
which agree well with the numerical results (see Fig.~\ref{fig:AII}).
For $N=5$, we obtain the representative moments as
\begin{align}
&\braket{r}
=
-\frac{651762902536369}{519519244124160} \nonumber \\
&\qquad\qquad+\frac{1488734661190876721\sqrt{3}\,\pi}
{4026368304825237504},
\\
&\braket{r^2}
=
\frac{7313533881933810221827}
{12079104914475712512000},
\\
&\braket{\cos \theta}
=
\frac{1387293479922263}
{45457933860864000}
\nonumber \\
&\qquad\qquad-\frac{9124207726819332547\sqrt{3}\,\pi}
{150988811430946406400},
\\
&\braket{\cos 2\theta} =
-\frac{88201737655469564356501}
{652271665381688475648000}.
\end{align}
For $N=6$, we obtain
\onecolumngrid
{
\begin{align}
&\braket{r}
=
-\frac{
8996088299729915875338415838941443932711593051947514379699
}{
5441188939509225262459593424896000000000000000000000000000
}
\nonumber\\
&\qquad\qquad\qquad\qquad\qquad\qquad
+\frac{
3015456452986538222806582215077179185643
}{
579606362761633055371055810550222028800
}
\arctan\frac{1}{2},
\\
&\braket{r^2}
=
\frac{
26230459934868728267727971870298987632189
}{
43149523156992000000000000000000000000000
},
\\
&\braket{\cos\theta}
=
\frac{
2077033152445253971103579917812659477134669076961933883
}{
340074308719326578903724589056000000000000000000000000000
}
\nonumber\\
&\qquad\qquad\qquad\qquad\qquad\qquad
-\frac{
7400574978116406687015793231339921553
}{
12075132557534021986896996053129625600
}
\arctan\frac{1}{2},
\\
&\braket{\cos 2\theta}
=
-\frac{
99778724481214336111395976293661664476711
}{
862990463139840000000000000000000000000000
}.
\end{align}
}
\twocolumngrid \noindent
For larger $N$, however, explicitly constructing $a_{N, l} \left( u, v \right)$ in Eq.~\eqref{eq:class-A2-general-algebraic} becomes computationally demanding.
Notably, although the $N=3$ distributions substantially deviate from the large-$N$ counterparts,
the $N=20$ distributions are already close to the large-$N$ results.
A characteristic feature of class AII$^{\dag}$ is its nonmonotonic finite-$N$ dependence.
For example, $\braket{r}$ initially increases with $N$ and subsequently decreases toward the large-$N$ numerical value,
which contrasts with the monotonic increase of $\braket{r}$ observed in classes A and AI$^{\dag}$.

\section{Class AI\texorpdfstring{$^{\dagger}$}{†}}
    \label{sec:AI}

\begin{figure}[t]
\centering
\includegraphics[width=1.0\linewidth]{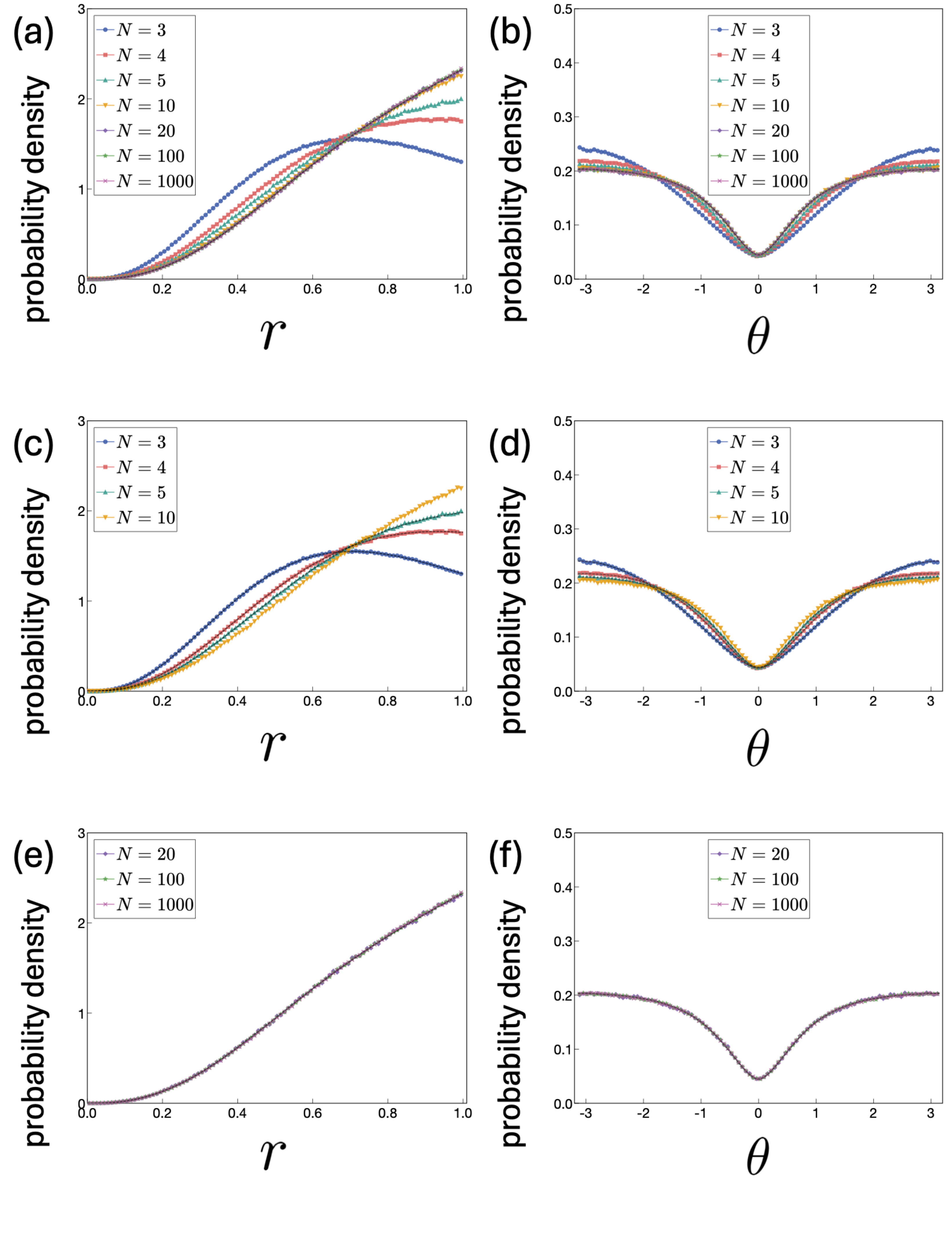} 
\caption{Probability densities of the modulus $r$ and argument $\theta$ of complex spacing ratios $re^{\ii \theta}$ for the Gaussian ensemble of non-Hermitian random matrices in class AI$^{\dag}$ with (a, b)~$N=3, 4, 5, 10, 20, 100, 1000$, (c, d)~$N=3, 4, 5, 10$, and (e, f)~$N=20, 100, 1000$.
The distributions are numerically obtained from $5 \times 10^6$ realizations for $N \leq 20$,
$10^5$ realizations for $N=100$, 
and $10^4$ realizations for $N=1000$.
Whereas the origin-conditioned distribution is considered for $N \leq 20$,
the central $50\%$ of the complex spectrum is employed for $N > 20$.
Black dashed curves:
analytical results in Eq.~\eqref{eq:p_exact_A1_N3} for $N=3$ and 
Monte Carlo estimates based on Eq.~\eqref{eq:MCMC} for $N=4, 5, 10, 20$.}	
    \label{fig:AI}
\end{figure}

In contrast to class AII$^{\dag}$,
the exact joint eigenvalue probability density for class AI$^{\dag}$ contains a noncompact orbital integral,
which makes the analytical evaluation more challenging.
Still, for $N=3$, we derive an exact integral representation of the complex-spacing-ratio distribution in Eq.~\eqref{eq:p_exact_A1_N3}.
For larger $N$, we confirm the consistency between the Monte Carlo estimates based on the exact joint eigenvalue probability density in Ref.~\cite{Xiao-26} and numerical results obtained from direct diagonalization of non-Hermitian random matrices, 
as shown in Fig.~\ref{fig:AI}.

\subsection{\texorpdfstring{$N=3$}{N=3}}

For $N=3$, the joint eigenvalue probability density function in class AI$^{\dag}$ reads~\cite{Xiao-26}
\begin{align}
    \rho_3 \left( z_1, z_2, z_3 \right) &= \frac{2}{3\pi^{7/2}} e^{- \sum_{n=1}^3 |z_n|^2} \left| \Delta_3 \right|^2 \mathcal{E} \left( a, b, c \right), \\
    \mathcal{E} \left( a, b, c \right) &\coloneqq \int_0^{\infty} ds \frac{e^{-s}}{\sqrt{\left( s+a \right) \left( s+b \right) \left( s+c \right)}} \nonumber \\
    &\quad \times K \left(  1 - \frac{abc}{\left( s+a \right) \left( s+b \right) \left( s+c \right)} \right)
        \label{eq:E}
\end{align}
with 
\begin{equation}
    a \coloneqq \left| z_1 - z_2 \right|^2,~~b \coloneqq \left| z_1 - z_3 \right|^2,~~c \coloneqq \left| z_2 - z_3 \right|^2.  
\end{equation}
Here, $K$ denotes the complete elliptic integral of the first kind:
\begin{equation}
    K \left( m \right) \coloneqq \int_0^{\pi/2} \frac{d\theta}{\sqrt{1-m\sin^2\theta}}.
\end{equation}
Then, we have $\left| \Delta_3 \right|^2 = abc = uvt^3$,
and Eq.~\eqref{eq:pN-t} becomes
\begin{equation}
    p_3^{(0)} \left( \eta \right) \propto uv \int_0^{\infty} dt\,t^4 e^{-(1+u)\,t} \mathcal{E} \left( ut, t, vt \right).
\end{equation}
In Eq.~\eqref{eq:E}, we introduce
\begin{equation}
    x \coloneqq \frac{s}{t},\quad Q \left( x \right) \coloneqq \left( x+1 \right) \left( x+u \right) \left( x+v \right),
\end{equation}
leading to
\begin{equation}
    \mathcal{E} \left( ut, t, vt \right) = \frac{1}{\sqrt{t}} \int_0^{\infty} dx \frac{e^{-tx}}{\sqrt{Q \left( x \right)}} K \left( 1 - \frac{uv}{Q \left( x \right)} \right),
\end{equation}
and hence
\begin{align}
    p_3^{(0)} \left( \eta \right) &\propto uv \int_0^{\infty} dx \frac{K \left( 1-uv/Q\left( x \right)\right)}{\sqrt{Q \left( x \right)}} \nonumber \\
    &\qquad\qquad \times \int_0^{\infty} dt\,t^{7/2} e^{-(x+1+u)\,t} \nonumber \\
    &\propto uv \int_0^{\infty} dx \frac{K \left( 1-uv/Q\left( x \right)\right)}{\sqrt{Q \left( x \right)} \left( x+1+u\right)^{9/2}}.
\end{align}
After imposing the normalization condition, we obtain
\begin{equation}
    p_3^{(0)} \left( \eta \right) = \frac{35uv}{2\pi} \int_0^{\infty} dx \frac{K \left( 1-uv/Q\left( x \right)\right)}{\sqrt{Q \left( x \right)} \left( x+1+u\right)^{9/2}}.
        \label{eq:p_exact_A1_N3}
\end{equation}
This expression provides an exact integral representation of the complex-spacing-ratio distribution for class AI$^{\dag}$ with $N=3$,
which agrees well with the numerical results, 
as shown in Fig.~\ref{fig:AI}.

From Eq.~\eqref{eq:p_exact_A1_N3}, we analytically extract the small-$r$ asymptotics.
For $0 < r \ll 1$, we have $u = r^2 \to 0$, $v = 1+r^2-2r\cos\theta \to 1$, and $Q \left( x \right) \to x \left( x+1 \right)^2$.
Using 
\begin{equation}
    K \left( 1-\varepsilon \right) = \log \frac{4}{\sqrt{\varepsilon}} + O \left( \varepsilon \log \frac{1}{\varepsilon} \right) \quad \left( 0 < \varepsilon \ll 1\right),
        \label{eq:elliptic_asymptotics}
\end{equation}
we also obtain for fixed $x>0$
\begin{equation}
    K \left( 1-\frac{uv}{Q \left( x \right)}\right) = \log \frac{1}{r} + O \left( 1 \right). 
\end{equation}
Consequently, the complex-spacing-ratio distribution exhibits the asymptotic behavior
\begin{align}    
    p_3^{(0)} \left( \eta \right) 
    &= \frac{35}{2\pi} r^2 \log \frac{1}{r} \int_0^{\infty} \frac{dx}{x^{1/2} \left( 1+x \right)^{11/2}} + O \left( r^2 \right) \nonumber \\
    &=\frac{128}{9\pi} r^2 \log \frac{1}{r} + O \left( r^2 \right),
        \label{eq:A1_N3_asymptotics_v1}
\end{align}
and the corresponding radial distribution satisfies
\begin{equation}
    p_{r, 3}^{(0)} \left( r \right) = \frac{256}{9} r^3 \log \frac{1}{r} + O \left( r^3 \right).
\end{equation}
The logarithmic correction $\log \left( 1/r \right)$ is a characteristic signature of class AI$^{\dag}$ that can also be found in the level-spacing distribution~\cite{Hamazaki-20, Xiao-26}.
The coefficient $256/9$ in $p_{r, 3}^{(0)} \left( r \right)$ is larger than the corresponding values $24$ for class A in Eq.~\eqref{eq:p_r_N3_A} and $112/9$ for class AII$^{\dag}$ in Eq.~\eqref{eq:p_r_N3_A2},
implying the weaker level repulsion induced by time-reversal symmetry$^{\dag}$.

We next derive the small-$\theta$ asymptotics from Eq.~\eqref{eq:p_exact_A1_N3}.
Let us instead consider the limit $\eta \to 1$,
for which we have $u \to 1$, $v \to 0$, and $Q \left( x \right) \to x \left( x+1 \right)^2$.
From Eq.~\eqref{eq:elliptic_asymptotics}, 
the elliptic integral $K \left( 1 - uv/Q \left( x \right) \right)$ in Eq.~\eqref{eq:p_exact_A1_N3} contains the logarithmically divergent term $\log 1/\sqrt{v}$.
Hence, Eq.~\eqref{eq:p_exact_A1_N3} can be asymptotically evaluated as
\begin{align}
    &p_3^{(0)} \left( \eta \right) = \frac{35v}{2\pi} \left[ \log \left( \frac{1}{\sqrt{v}} \right) \right. \nonumber \\
    &\qquad\quad \left. \times\int_0^{\infty} \frac{dx}{\sqrt{x} \left( x+1 \right) \left( x+2\right)^{9/2}} + O \left( 1 \right)\right] \nonumber \\
    &\qquad= D_0 \left| 1-\eta \right|^2 \log \frac{1}{\left| 1-\eta \right|} + O\,( \left| 1-\eta \right|^2 ),
\end{align}
with 
\begin{align}
    D_0 &\coloneqq \frac{35}{2\pi} \int_0^{\infty} \frac{dx}{\sqrt{x} \left( x+1 \right) \left( x+2\right)^{9/2}} \nonumber \\
    &= \frac{35}{4} - \frac{80}{3\pi} 
    = 0.261736 \cdots.
\end{align}
The level repulsion here arises from the collision between $z_2$ and $z_3$,
whereas that in Eq.~\eqref{eq:A1_N3_asymptotics_v1} arises from the collision between $z_1$ and $z_2$.

The logarithmic singularity in $p_3^{(0)} \left( \eta \right)$ gives rise to nonanalyticity in the angular marginal density $p_{\theta, 3}^{(0)} \left( \theta \right)$ in Eq.~\eqref{eq:marginal_theta}.
Since the $\theta$-dependent singularity in $p_3^{(0)} \left( \eta \right)$ occurs only for $\eta \to 1$,
the leading singular contribution to $p_{\theta, 3}^{(0)} \left( \theta \right)$ is governed by the region around $r=1$.
In the following, we focus on this singular contribution.
Introducing $\varepsilon \coloneqq 1-r$ ($0 < \varepsilon \ll 1$),
we have
\begin{equation}
    v = \left| 1-\eta \right|^2 = \left( 1-r \right)^2 + 2r \left( 1- \cos \theta \right) \simeq \varepsilon^2 + \theta^2,
\end{equation}
and hence
\begin{equation}
    p_{\theta, 3}^{(0)} \left( \theta \right) \simeq - \frac{D_0}{2} \int_0^{\varepsilon_0} d\varepsilon \left( \varepsilon^2 + \theta^2 \right) \log \left( \varepsilon^2 + \theta^2 \right)
\end{equation}
with a cutoff $\varepsilon_0 > 0$.
It follows that
\begin{align}
    \frac{d^3 p_{\theta, 3}^{(0)}}{d\theta^3} \left( \theta \right) &= - 2D_0 \int_0^{\varepsilon_0} d\varepsilon \frac{\theta \left( 3 \varepsilon^2 + \theta^2 \right)}{\left( \varepsilon^2 + \theta^2 \right)^2} \nonumber \\
    &= - 2D_0 \int_0^{\varepsilon_0/\theta} dy \frac{3y^2+1}{\left( y^2+1 \right)^2}
\end{align}
with $y \coloneqq \varepsilon/\theta$.
Hence, we have the discontinuous behavior at $\theta = 0$,
\begin{equation}
    \frac{d^3 p_{\theta, 3}^{(0)}}{d\theta^3} \left( \theta \right) \to \begin{cases}
        -2\pi D_0 & \left( \theta \to 0^+ \right), \\
        +2\pi D_0 & \left( \theta \to 0^- \right), \\
    \end{cases}
\end{equation}
where we use 
\begin{equation}
    \int_0^{\infty} dy \frac{3y^2 + 1}{\left( y^2 + 1\right)^2} = \pi.
\end{equation}
Consequently, we have for $\theta \to 0$
\begin{equation}
    p_{\theta, 3}^{(0)} \left( \theta \right) = p_{\theta, 3}^{(0)} \left( 0 \right) + A_2 \theta^2 - \frac{\pi D_0}{3} \left| \theta \right|^3 + o\,( \left| \theta \right|^3 )
\end{equation}
with a constant $A_2 > 0$.
Notably, $p_{\theta, 3}^{(0)}$ contains the nonanalytic term $\left| \theta \right|^3$,
in sharp contrast with class A in Eq.~\eqref{eq:p_theta_N3_A} and class AII$^{\dag}$ in Eq.~\eqref{eq:p_theta_N3_A2}.

\begin{figure}[t]
\centering
\includegraphics[width=0.7\linewidth]{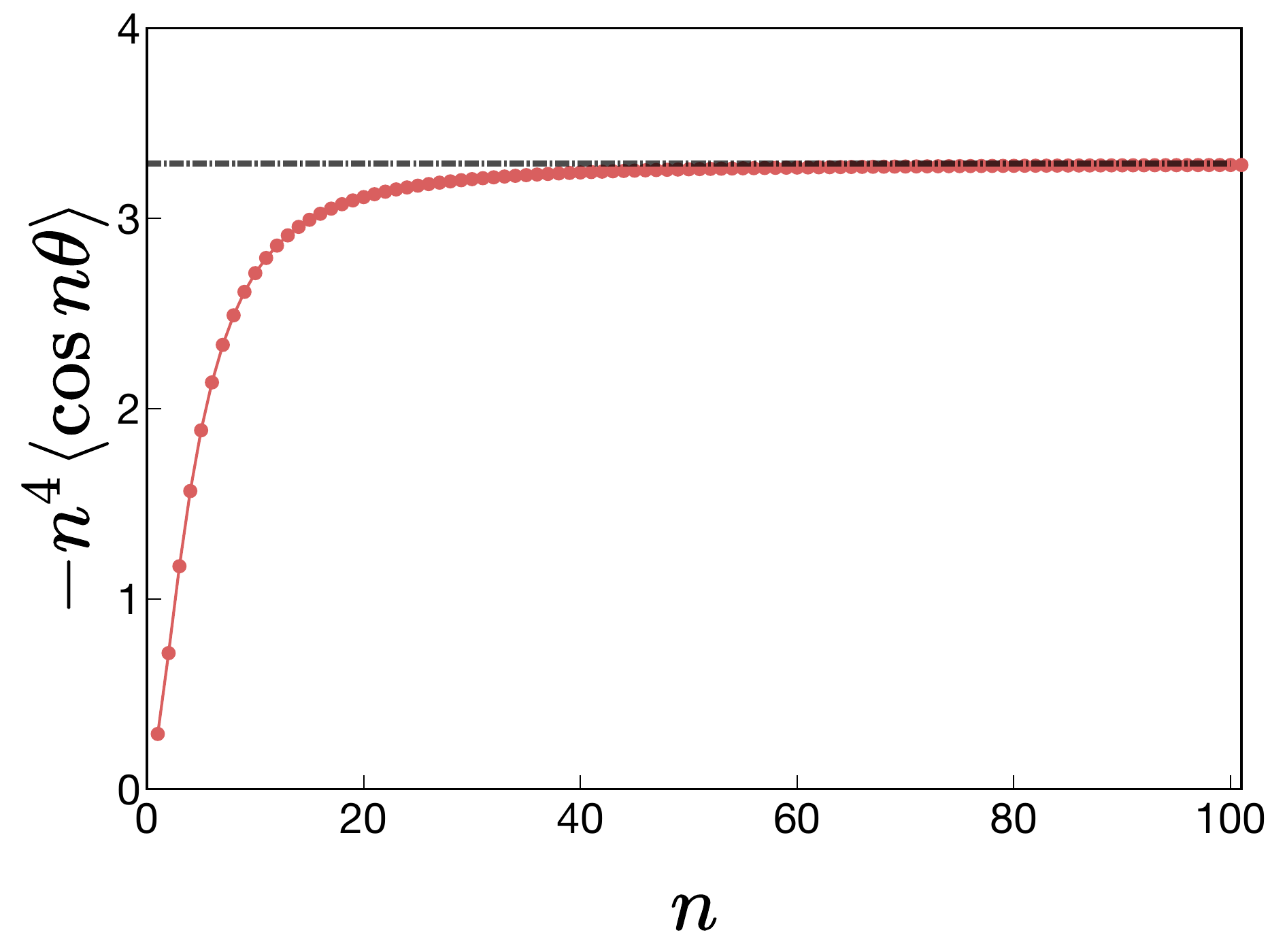} 
\caption{Asymptotic angular moments of complex spacing ratios for $N=3$ non-Hermitian random matrices in class AI$^{\dag}$.
The quantity $-n^4 \braket{\cos n \theta}$ is plotted as a function of $n$.
Red dots: analytical results in Eq.~\eqref{eq:p_exact_A1_N3}.
Black dashed line: asymptotic value in Eq.~\eqref{eq:A1_n^4}.}	
    \label{fig:AI_asymptotic}
\end{figure}

Moreover, we have
\begin{equation}
    \frac{d^4 p_{\theta, 3}^{(0)}}{d\theta^4} \left( \theta \right) \simeq -4\pi D_0 \delta \left( \theta \right) \quad \left( \left| \theta \right| \ll 1 \right),
        \label{eq:A1_p3_delta}
\end{equation}
yielding
\begin{align}
    n^4 \braket{\cos n\theta} &= \int_{-\pi}^{\pi} d\theta\,p_{\theta, 3}^{(0)} \left( \theta \right) \left( \frac{d^4}{d\theta^4} \cos n\theta \right) \nonumber \\
    &= \int_{-\pi}^{\pi} d\theta \left( \frac{d^4 p_{\theta, 3}^{(0)}}{d\theta^4} \left( \theta \right) \right) \cos n\theta \nonumber \\
    &\simeq -4\pi D_0.
\end{align}
Therefore, the asymptotic angular moments are  
\begin{equation}
    \braket{\cos n\theta} \simeq - \frac{4\pi D_0}{n^4} = - \frac{3.28908\cdots}{n^4}.
        \label{eq:A1_n^4}
\end{equation}
Importantly, this result shows that $\braket{\cos n\theta}$ remains nonvanishing even for arbitrarily large $n$ in class AI$^{\dag}$,
whereas it vanishes for $n \geq 2$ in class A and $n \geq 3$ in class AII$^{\dag}$ [see Eqs.~\eqref{eq:p_theta_N3_A} and \eqref{eq:p_theta_N3_A2}].
The persistence of these higher-order angular harmonics represents another hallmark of class AI$^{\dag}$.
It should be noted, however, that Eq.~\eqref{eq:A1_p3_delta} can generally contain additional regular contributions,
which can lead to corrections to Eq.~\eqref{eq:A1_n^4}.
Still, we confirm Eq.~\eqref{eq:A1_n^4} for large $n$, as shown in Fig.~\ref{fig:AI_asymptotic}.

\subsection{Arbitrary \texorpdfstring{$N$}{N}}

For generic $N$, the joint eigenvalue probability density contains a noncompact orbital integral~\cite{Xiao-26}.
Instead of evaluating this integral directly,
we here sample complex eigenvalues and obtain the complex-spacing-ratio distribution,
employing the Monte Carlo method with respect to the measure
\begin{align}
    &d\mu_N \propto \left| \Delta_N \left( 0, z_2, \cdots, z_N \right) \right|^2 j_N \left( B \right) e^{-\bm{z}^{\dag} T \left( B \right) \bm{z}} \nonumber \\
    &\qquad\qquad\qquad\qquad\times \prod_{i<j} dB_{ij} \prod_{n=2}^{N} d^2 z_n
        \label{eq:MCMC}
\end{align}
under the condition $z_1 = 0$ (see Appendix~\ref{appendix:A1} for details).
This procedure is analogous to the Monte Carlo evaluation of complex spacing distributions in Ref.~\cite{Xiao-26}.
Importantly, this measure follows directly from the exact joint probability density derived in Ref.~\cite{Xiao-26} and does not rely on numerically generating and diagonalizing non-Hermitian random matrices.
Using this measure, we obtain the complex-spacing-ratio distributions and associated moments for $N=4, 5, 10, 20$, 
as summarized in Table~\ref{tab:analytical} and Fig.~\ref{fig:AI}.
These results are consistent with those obtained by direct numerical diagonalization.

\section{Conclusions}
    \label{sec:conclusion}

Our results connect the many-eigenvalue correlations of classes AI$^{\dag}$ and AII$^{\dag}$ reported in Ref.~\cite{Xiao-26} to the radial and angular statistics of complex spacing ratios. 
Retaining the neighbor-ordering constraints provides a finite-size description beyond characterization based solely on the leading power law of level repulsion. 
The resulting origin-conditioned distributions serve as quantitative benchmarks for the corresponding Gaussian ensembles.
In class AII$^{\dag}$,
the nonmonotonic size dependence suggests the importance of distinguishing small-$N$ behavior from universal bulk statistics.
In class AI$^{\dag}$,
the logarithmic collision law gives rise to a nonanalytic angular density and algebraically decaying Fourier harmonics for $N=3$,
which contrast with the finite harmonic
content for classes A and AII$^{\dag}$.

Deriving the universal complex-spacing-ratio distributions in the large-$N$ limit remains an important open problem,
especially for classes AI$^{\dag}$ and AII$^{\dag}$. 
The finite-size algebraic construction for class AII$^{\dag}$ and the integral representation for class AI$^{\dag}$ provide starting points for a controlled asymptotic analysis.
A related challenge is to establish the conditions under which the limiting distributions become insensitive to the choice of ensemble. 
Extending the analysis beyond the Gaussian ensembles while preserving the symmetry constraints should distinguish universal correlations from ensemble-dependent finite-size effects. 
Studying the large-$N$ distributions and their ensemble dependence should clarify the applicability of the non-Hermitian threefold way and provide a basis for using complex spacing ratios to characterize universality in the physics of open quantum systems.

\medskip
\begingroup
\renewcommand{\addcontentsline}[3]{}
\begin{acknowledgments}
K.K. is supported by JSPS KAKENHI Grants 
No.~JP26H02015, No.~JP26K06970, and No.~JP26K17046, 
and JST FOREST Program Grant No.~JPMJFR256P.
K.K. thanks the Yukawa Institute for Theoretical Physics at Kyoto University,
where this work was partially completed during the workshop ``Localisation 2026" (YITP-W-26-10).
OpenAI's ChatGPT (GPT-6 Astra) was used to assist with code development and debugging, and to check selected analytical steps. 
All outputs were critically reviewed and independently verified through analytical derivations and numerical tests.
\end{acknowledgments}
\endgroup

\appendix

\section{Numerical calculations of origin-conditioned complex-spacing-ratio distributions}
    \label{appendix:origin}

To obtain the results for $N \leq 20$ in Table~\ref{tab:numerical},
we numerically calculate the origin-conditioned complex-spacing-ratio distributions in Eq.~\eqref{eq:origin-conditioned}.
We decompose a numerically generated non-Hermitian random matrix $H$ as
\begin{equation}
    H = H_0 + c I, \quad \mathrm{Tr}\,H_0 = 0,
\end{equation}
where
\begin{equation}
    c = \begin{cases}
        \mathrm{Tr}\,H/N & ( \text{class~AI}^{\dag} ), \\
        \mathrm{Tr}\,H/2N & ( \text{class~AII}^{\dag} ), \\
    \end{cases}
\end{equation}
is distributed according to 
\begin{equation}
    p \left( c \right) = \frac{N}{\pi} e^{-N |c|^2}.
\end{equation}
Let $z_1$, $\cdots$, $z_N$ denote different complex eigenvalues of $H$ and $\xi_1$, $\cdots$, $\xi_N$ denote the corresponding eigenvalues of $H_0$,
satisfying $z_n = \xi_n + c$.
For given $\xi_n$ and its nearest neighbor $\xi_{{\rm NN} (n)}$ and next-to-nearest neighbor $\xi_{{\rm NNN} (n)}$,
the complex spacing ratio in Eq.~\eqref{eq:CSR} is given as
\begin{equation}
    \eta_n = \frac{\xi_{{\rm NN} (n)} - \xi_n}{\xi_{{\rm NNN} (n)} - \xi_n}.
\end{equation}
Conditioning the reference eigenvalue $z_n$ to be zero is equivalent to imposing $c = -\xi_n$.
Consequently, moments of $\eta$ with respect to the origin-conditioned probability distribution are given as
\begin{equation}
    \braket{f \left( \eta \right)}^{(0)} = \frac{\mathbb{E}_{H_0} \left[ \sum_{n=1}^{N} e^{-N |\xi_n|^2} f \left( \eta_n \right) \right]}{\mathbb{E}_{H_0} \left[ \sum_{n=1}^{N} e^{-N |\xi_n|^2} \right]}
        \label{aeq:conditioned_zero}
\end{equation}
for an arbitrary function $f$.

The trace-free matrix $H_0$ can further be decomposed into
\begin{equation}
    H_0 = \sqrt{s} \hat{H}_0,
\end{equation}
where the radial scale
\begin{equation}
    s = \begin{cases}
        \mathrm{Tr}\,H_0^{\dag} H_0 & ( \text{class~AI}^{\dag} ), \\
        \mathrm{Tr}\,H_0^{\dag} H_0/2 & ( \text{class~AII}^{\dag} ), \\
    \end{cases}
\end{equation}
is distributed according to 
\begin{equation}
    p \left( s \right) = \frac{s^{d-1} e^{-s}}{\Gamma \left( d \right)},
\end{equation}
with 
\begin{equation}
d = \begin{cases}
        N \left( N+1 \right)/2 - 1 & ( \text{class~AI}^{\dag} ), \\
        N \left( 2N-1 \right) - 1 & ( \text{class~AII}^{\dag} ). \\
\end{cases}
\end{equation}
Further introducing 
\begin{equation}
    \hat{\xi}_n \coloneqq \frac{\xi_n}{\sqrt{s}},
\end{equation}
we can analytically integrate out the radial scale $s$ by
\begin{align}
    w_n &\coloneqq \int_0^{\infty} ds' \frac{(s')^{d-1} e^{-s'}}{\Gamma \left( d \right)} e^{-Ns' |\hat{\xi}_n|^2} \nonumber \\
    &= \frac{1}{( 1 + N\,| \hat{\xi}_n |^2 )^{d}},
\end{align}
reducing Eq.~\eqref{aeq:conditioned_zero} to
\begin{equation}
    \braket{f \left( \eta \right)}^{(0)} = \frac{\mathbb{E}_{H_0} \left[ \sum_{n=1}^{N} w_n f \left( \eta_n \right) \right]}{\mathbb{E}_{H_0} \left[ \sum_{n=1}^{N} w_n \right]}.
\end{equation}
    
\section{Class AI\texorpdfstring{$^{\dagger}$}{†} for arbitrary \texorpdfstring{$N$}{N}}
    \label{appendix:A1}

For generic $N$,
non-Hermitian matrices $H$ in class AI$^{\dag}$,
satisfying $H^T = H$,
can be diagonalized as
\begin{equation}
    H = O\,\mathrm{diag} \left( z_1, \cdots, z_N \right) O^T, \quad O^T O = I_N,
\end{equation}
where the eigenvalue spectrum is assumed to be nondegenerate.
The complex orthogonal matrix $O$ can be parametrized by
\begin{equation}
    O = U e^{\ii B},~~U \in \mathrm{SO} \left( N \right),~~B^T = -B,~~B \in \mathbb{R}^{N\times N}.
\end{equation}
Here, for fixed eigenvalues, 
$B$ effectively captures the nonunitary degrees of freedom associated with eigenvectors.
Then, the joint eigenvalue probability density function reads~\cite{Xiao-26}
\begin{align}
    &\rho_N \left( z_1, \cdots, z_N \right) = K_N \left| \Delta_N \left( z_1, \cdots, z_N \right) \right|^2 \nonumber \\
    &\qquad \times \int_{\mathbb{R}^{N(N-1)/2}} \left( \prod_{i<j} dB_{ij} \right) j_N \left( B \right) e^{-\bm{z}^{\dag} T \left( B \right) \bm{z}}
\end{align}
with 
\begin{equation}
    K_N \coloneqq \frac{2^{N(N-1)/2}}{N! \pi^{(N^2+N+2)/4} \prod_{j=2}^{N} \Gamma \left( j/2\right)}.
\end{equation}
In contrast to class AII$^{\dag}$,
the integral over the eigenvector degrees of freedom remains, with
\begin{equation}
    T_{ij} \left( B \right) \coloneqq \left[ \left( e^{2\ii B}\right)_{ij} \right]^2,
\end{equation}
and 
\begin{align}
    &j_N \left( B \right) \coloneqq \prod_{a<b} \left[ \frac{\sinh \left( \sigma_a + \sigma_b \right)}{\sigma_a + \sigma_b} \frac{\sinh \left( \sigma_a - \sigma_b \right)}{\sigma_a - \sigma_b} \right]^2 \nonumber \\
    &\qquad\qquad\qquad \times \begin{cases}
        \displaystyle
        \prod_{a} \left( \cfrac{\sinh \sigma_a}{\sigma_a}\right)^2 & \left( \text{odd}~N \right), \\
        1 & \left( \text{even}~N \right),
    \end{cases}
\end{align}
where nonzero eigenvalues of $B$ correspond to $\pm \ii \sigma_a$.

Introducing
\begin{equation}
    \bm{\xi} \coloneqq \left( 0, \eta, 1, \xi_4, \cdots, \xi_N \right)^T, \quad \xi_n \coloneqq \frac{z_n}{\sqrt{t}},
\end{equation}
we have $\left| \Delta_N\,( \sqrt{t} \bm{\xi} ) \right|^2 = t^{N(N-1)/2} \left| \Delta_N \left( \bm{\xi} \right) \right|^2$ and $\prod_{n=4}^{N} d^2z_n = t^{N-3} \prod_{n=4}^{N} d^2\xi_n$.
Then, the integral over $t$ in Eq.~\eqref{eq:pN-t} is 
\begin{align}
    &\int_0^{\infty} dt\,t^{N(N+1)/2-2} e^{-t \bm{\xi}^{\dag} T \left( B \right) \bm{\xi}} \nonumber \\
    &\qquad\quad = \frac{\Gamma \left( N(N+1)/2-1 \right)}{\left( \bm{\xi}^{\dag} T \left( B \right) \bm{\xi} \right)^{N(N+1)/2-1}},
\end{align}
leading to
\begin{align}
    &p_N^{(0)} \left( \eta \right)
    = \frac{1}{2} \left( N-1 \right) \left( N-2 \right) \left( N+1 \right) \pi^2 K_N \nonumber \\
    &~\times \Gamma \left( N(N+1)/2-1 \right) \int_{\mathbb{R}^{N(N-1)/2}} \left( \prod_{i<j} dB_{ij} \right) j_N \left( B \right) \nonumber \\
    &~\times \int_{\left| \xi_n \right| > 1} \left( \prod_{n=4}^{N} d^2\xi_n \right) \frac{\left| \Delta_N \left( \bm{\xi} \right) \right|^2}{\left( \bm{\xi}^{\dag} T \left( B \right) \bm{\xi} \right)^{N(N+1)/2-1}}.
\end{align}

\let\oldaddcontentsline\addcontentsline
\renewcommand{\addcontentsline}[3]{}
\bibliography{ref.bib}
\let\addcontentsline\oldaddcontentsline

\end{document}